\documentclass[12pt]{article}
\usepackage{amsmath,amssymb,bbm,stmaryrd}

\def\be{\begin{equation}}
\def\ee{\end{equation}}
\def\bea{\begin{eqnarray}}
\def\eea{\end{eqnarray}}

\newcommand{\pa}{\partial}
\newcommand{\ft}[2]{{\textstyle\frac{#1}{#2}}}
\newcommand{\nn}{\nonumber}

\def\slashchar#1{\setbox0=\hbox{$#1$}           
   \dimen0=\wd0                                 
   \setbox1=\hbox{/} \dimen1=\wd1               
   \ifdim\dimen0>\dimen1                        
      \rlap{\hbox to \dimen0{\hfil/\hfil}}      
      #1                                        
   \else                                        
      \rlap{\hbox to \dimen1{\hfil$#1$\hfil}}   
      /                                         
   \fi}

\def\diag{\text{diag}}

\def\mua{\mu}
\def\mub{\nu}
\def\muc{\rho}
\def\mud{\lambda}
\def\mue{\sigma}
\def\muf{\tau}
\def\mug{\kappa}
\def\ma{m}
\def\mb{n}
\def\mc{p}
\def\md{q}
\def\me{r}

\def\Ja{M}
\def\Jb{N}
\def\Jc{P}
\def\Jd{Q}
\def\Je{R}
\def\Jf{S}
\def\Jg{T}
\def\Jh{U}
\def\Ji{V}
\def\Jj{W}
\def\Jk{X}
\def\Jl{Y}
\def\Jm{Z}
\def\Jn{A}
\def\Jo{B}
\def\Jp{C}
\def\Jq{D}
\def\Jr{E}
\def\ja{a}
\def\jb{b}
\def\jc{c}
\def\jd{d}
\def\je{e}
\def\jf{f}
\def\jg{g}
\def\jh{h}
\def\ji{i}
\def\jj{j}
\def\jk{k}
\def\jl{l}
\def\jm{m}
\def\jn{n}
\def\ka{\alpha}
\def\kb{\beta}
\def\kc{\gamma}
\def\kd{\delta}

\def\la{x}

\newdimen\squaresize \squaresize=12pt
\newdimen\thickness \thickness=0.7pt

\def\square#1{\hbox{\vrule width \thickness
   \vbox to \squaresize{\hrule height \thickness\vss
      \hbox to \squaresize{\hss#1\hss}
   \vss\hrule height\thickness}
\unskip\vrule width \thickness} \kern-\thickness}

\def\cut#1{\hbox{\vrule width-1 \thickness
   \vbox to \squaresize{\hrule height-1 \thickness\vss
      \hbox to \squaresize{\hss#1\hss}
   \vss\hrule height-1\thickness}
\unskip\vrule width +4 \thickness} \kern-\thickness}

\def\vsquare#1{\vbox{\square{$#1$}}\kern-\thickness}

\def\young#1{
\vbox{\smallskip\offinterlineskip \halign{&\vsquare{##}\cr #1}}}

\newcommand{\tinyyoung}[1]{
\squaresize=7pt \thickness=0.4pt \mbox{\tiny\young{#1}}
\squaresize=12pt \thickness=0.7pt}

\begin{document}

\begin{titlepage}
\begin{center}

\hfill {\tt DESY/05-101}\\
\hfill {\tt ZMP-HH/05-010}

\vskip 1.5cm 
\begin{center}
{\Large {\bf THE MAXIMAL D$=$7 SUPERGRAVITIES}}
\end{center}

\vskip 1.0cm

{\bf Henning Samtleben and Martin Weidner} \\

\vskip 22pt

{\em II. Institut f\"ur Theoretische Physik\\[-.6ex]
Universit\"at Hamburg\\[-.6ex]
Luruper Chaussee 149\\[-.6ex]
D-22761 Hamburg, Germany}

\vskip 8pt
and
\vskip 8pt

{\em Zentrum f\"ur Mathematische Physik\\[-.6ex]
Universit\"at Hamburg\\[-.6ex]
Bundesstrasse 55\\[-.6ex]
D-20146 Hamburg, Germany}\\

\vskip 15pt

{{\tt henning.samtleben@desy.de, martin.weidner@desy.de}} \\

\vskip 0.8cm

\end{center}

\vskip 12mm

\begin{center} {\bf ABSTRACT}\\[3ex]

\begin{minipage}{13cm}
\small

The general seven-dimensional maximal supergravity is presented.
Its universal Lagrangian is described in terms of an embedding tensor 
which can be characterized group-theoretically. 
The theory generically combines vector, two-form and three-form tensor fields that 
transform into each other under an intricate set of nonabelian gauge transformations.
The embedding tensor encodes the proper distribution of 
the degrees of freedom among these fields.
In addition to the kinetic terms the vector and tensor fields 
contribute to the Lagrangian with a unique gauge invariant 
Chern-Simons term. 
This new formulation encompasses all possible gaugings. 
Examples include the sphere reductions of M theory and of  the type IIA/IIB theories
with gauge groups ${\rm SO}(5)$, ${\rm CSO}(4,1)$, and ${\rm SO}(4)$,
respectively.

\end{minipage}
\end{center}
\noindent

\vfill

June 2005

\end{titlepage}

\tableofcontents

\newpage

\section{Introduction}
\setcounter{equation}{0}

Over recent years it has emerged that the structures of supergravity theories 
with a maximal number of supercharges are far richer than originally
anticipated~\cite{Nicolai:2000sc,deWit:2002vt,deWit:2004nw,deWit:2005hv}.
Maximal supergravities are obtained as deformations (gaugings) of 
toroidally compactified eleven-dimensional supergravity
by coupling the initially abelian vector fields to charges assigned to the elementary fields.
Two features have proven universal
in the construction of these theories. 
First, in dimension $D=11-d$ 
it is the global symmetry group ${\rm G}={\rm E}_{d(d)}$ 
of the toroidally compactified theory 
which not only organizes the ungauged theory
but also its possible deformations.
The gaugings are parametrized in terms of a constant 
embedding tensor $\Theta$.
When treating this embedding tensor as a spurionic object
that transforms under ${\rm G}$, the Lagrangian and transformation rules
remain formally invariant under ${\rm G}$.
Consistency of the theory can then be encoded in
a number of representation constraints on $\Theta$.

Second, the gaugings 
generically involve $p$-form tensor fields together 
with their dual $D-p-2$ forms.
For the ungauged theory it is known that in order
to exhibit the full global symmetry group ${\rm G}$
all tensor fields have to be dualized
into forms of lowest possible rank ---
employing the on-shell duality between
antisymmetric tensor fields of rank $p$ and of rank $D-p-2$. 
In contrast, the generic gauging combines
$p$-form fields together with their duals
which come in mutually conjugate ${\rm G}$ representations.
The specific form of the embedding tensor 
in a particular gauging encodes the proper distribution 
of the degrees of freedom among these fields.
Together, this gave rise to a universal formulation
of the maximal supergravities in various space-time 
dimensions~\cite{Nicolai:2000sc,deWit:2002vt,deWit:2004nw,deWit:2005hv},
capturing all possible supersymmetric deformations
in a manifestly ${\rm G}$-covariant way.
Although most of this formalism has been established for 
the global symmetry groups ${\rm G}={\rm E}_{d(d)}$ 
of the maximal supergravities the structures are not restricted
to maximal supersymmetry and similarly underlay
the theories with lower number of supercharges.

In this paper we realize this program for the maximal $D=7$ case.
The ungauged maximal supergravity in seven dimensions 
possesses a global
${\rm E}_{4(4)}={\rm SL}(5)$ symmetry~\cite{Sezgin:1982gi}.
This theory is formulated entirely in terms of vector and two-form
tensor fields, transforming in the ${\bf\overline{10}}$
and the ${\bf 5}$ representation of ${\rm SL}(5)$, respectively.
The first gaugings in $D=7$ were constructed in
\cite{Pernici:1984xx,Pernici:1984zw}
with semisimple gauge groups ${\rm SO}(5)$, ${\rm SO}(4,1)$, and ${\rm SO}(3,2)$.
Notably, the ${\rm SO}(5)$ theory corresponds to compactification of 
$D=11$ supergravity on the sphere~$S^{4}$.
Instead of the two-forms these theories feature five massive selfdual three-forms. 
Selfduality ensures that the three-forms carry the same number of degrees of 
freedom as massless two-forms and thus the total number
of degrees of freedom is unchanged as required by supersymmetry~\cite{Townsend:1983xs}.
Global ${\rm SL}(5)$ invariance is manifestly broken in these theories.
That this is not the full story can readily be deduced from the fact
that for instance none of these gaugings describes the maximal theories 
expected to descend from the ten-dimensional type IIA/IIB theories by compactification 
on a three-sphere~$S^{3}$. Indeed, in~\cite{Cvetic:2000ah} the bosonic part
of a theory with non-semisimple gauge group ${\rm CSO}(4,0,1)$ was constructed
and shown to describe the (warped) $S^{3}$ compactification of type IIA supergravity.
This theory combines a single massless two-form with four massive selfdual three-forms,
thus giving rise to yet another distribution of the degrees of freedom.
Other theories, such as the maximal ${\rm SO}(4)$ gauging expected from the 
type IIB reduction with only two-form tensor fields in the spectrum had not yet even been 
constructed.
Different gaugings in seven dimensions thus seem to appear with different
field representations according to how many of the two-form tensor fields
have been dualized into three-form tensors. This circumstance together
with the fact that every dualization appears to manifestly break the global
${\rm SL}(5)$ symmetry of the ungauged theory, has hampered a systematic
analysis of the seven-dimensional gaugings.

The formalism we will adopt in this paper in contrast 
is flexible enough to comprise all different gaugings 
in a single universal formulation. In accordance with the
general scheme explained above it employs vector fields and 
two-form tensor fields together with three-form tensor fields 
transforming in the ${\bf\overline{5}}$ representation of ${\rm SL}(5)$.
Duality between two-form and three-form tensor fields
arises as an equation of motion from the universal Lagrangian.
The gauging is entirely parametrized by means of a constant {embedding
tensor} $\Theta$ which carries the structure of a ${\bf 15}+{\bf\overline{40}}$ 
representation of ${\rm SL}(5)$ and describes the embedding
of the gauge group ${\rm G}_{0}$ into ${\rm SL}(5)$.
When the embedding tensor transforms according to this representation,
the full Lagrangian and transformation rules remain formally ${\rm SL}(5)$
invariant. Only after freezing the embedding tensor to a constant, i.e.\
choosing a particular gauging, the global symmetry is broken down to 
the gauge group~${\rm G}_{0}\subset{\rm SL}(5)$.

The embedding tensor describes the minimal couplings of vectors to scalars
while at the same time its components in the 
${\bf\overline{40}}$ and the ${\bf 15}$ representation are
precisely tailored such as to introduce additional St\"uckelberg type 
couplings between vector and two-form tensors
and between two-form and three-form tensors, respectively.
Altogether, the embedding tensor defines a set of nonabelian 
gauge transformations between vector and tensor fields which ensures
that the full system always describes 100 degrees of freedom as required 
by maximal supersymmetry. The precise form of a given embedding tensor determines
which fields actually participate in the particular gauging and how the degrees of freedom
are distributed among them.
As particular applications of this universal formulation we recover
the known seven-dimensional 
gaugings as well as a number of new examples.
In particular, we obtain the maximal theory with compact gauge 
group ${\rm SO}(4)$ that is expected to describe the (warped)
$S^{3}$ reduction of type IIB supergravity.

This paper is organized as follows. In section~\ref{sec:embedding}
we introduce the embedding tensor for seven-dimensional maximal
supergravity and discuss its ${\rm SL}(5)$ representation constraints and their
consequences. 
In particular, the embedding tensor parametrizes the extension of 
the abelian vector/tensor system of the ungauged theory to a 
system combining vector, two-form and three-form tensor fields
and their nonabelian gauge invariances. This is 
presented in section~\ref{SecVecTen} together with the possible
gauge invariant couplings in seven dimensions, in particular
a novel Chern-Simons type term involving all these fields.
In section~\ref{sec:CT} we discuss properties of the scalar coset 
space ${\rm SL}(5)/{\rm SO}(5)$ and define the so-called $T$-tensor
which is naturally derived from the embedding tensor and encodes the extra 
coupling of the scalars to the fermions that are added 
in the process of gauging.
The main results of this paper are presented in section~\ref{SecLagrSUSY}
where we give the universal seven-dimensional Lagrangian and
the supersymmetry transformation rules, both parametrized
in terms of the embedding tensor.
Finally, in section \ref{SecExpamples} we 
illustrate the general formalism with a number of representative examples. 
Some technical details of the computations are relegated to three appendices.

\section{The embedding tensor}
\setcounter{equation}{0}
\label{sec:embedding}

The global symmetry group of the ungauged seven-dimensional theory 
is $E_{4(4)}={\rm SL}(5)$. Its 24 generators $t^{\Ja}{}\!_{\Jb}$ are labeled
by indices $\Ja, \Jb=1, \dots, 5$ with $t^{\Ja}{}\!_{\Ja}=0$ and satisfy
the algebra
\bea
\Big[\,t^{\Ja}{}\!_{\Jb}\,,\;t^{\Jc}{}\!_{\Jd}\,\Big] &=&
\delta^{\Jc}_{\Jb}\,t^{\Ja}{}\!^{\phantom{\Jc}}_{\Jd}
-
\delta^{\Ja}_{\Jd}\,t^{\Jc}{}\!^{\phantom{\Jc}}_{\Jb} 
\;.
\label{sl5}
\eea
The (abelian) vector fields $A_\mu^{\vphantom{[]}\Ja\Jb}=A_\mu^{[\Ja\Jb]}$ 
of the ungauged theory transform in the representation 
$\overline{\bf 10}$ of ${\rm SL}(5)$, so that
$\delta A_\mu^{\vphantom{[]}\Ja\Jb}=
2\Lambda_{\Jc}{}_{\vphantom{[]}}^{[\Ja}A_{\mu}^{\vphantom{[]}\Jb]\Jc}$.
The two-form tensor fields $B_{\mu\nu\,\Ja}$
 transform in the ${\bf 5}\,$ representation.

A gauging is encoded in a real {\it embedding tensor}\, 
$\Theta_{\Ja\Jb,\Jc}{}^{\Jd}=\Theta_{[\Ja\Jb],\Jc}{}^{\Jd}$
which identifies the generators $X_{\Ja\Jb}=X_{[\Ja\Jb]}$ of the gauge group ${\rm G}_{0}$
among the ${\rm SL}(5)$ generators
according to
\bea
\label{X-theta}
X_{\Ja\Jb} &=& \Theta_{\Ja\Jb,\Jc}{}^{\Jd}\;
t^{\Jc}{}\!_{\Jd}\;.
\eea
It acts as a projector whose rank equals the dimension of the 
gauge group
up to central extensions. Covariant derivatives
take the form
\bea
D_{\mu} &=& \nabla_{\mu} 
- g A_{\mu}^{\vphantom{[]}\Ja\Jb}\Theta_{\Ja\Jb,\Jc}{}^{\Jd}\,t^{\Jc}{}\!_{\Jd}
\;,
\eea
where we have introduced the gauge coupling constant $g$.
In our construction we will treat the embedding tensor as a spurionic
object that transforms under ${\rm SL}(5)$, so that the Lagrangian and 
transformation rules remain formally ${\rm SL}(5)$ covariant. The embedding
tensor can then be characterized group-theoretically. When freezing
$\Theta_{\Ja\Jb,\Jc}{}^{\Jd}$ to a constant, the ${\rm SL}(5)$-invariance is broken. 
It has emerged in the recent studies of maximal supergravity theories
that consistency of the gauging is typically encoded in a set of 
representation constraints on the embedding 
tensor~\cite{Nicolai:2000sc, deWit:2002vt, deWit:2004nw}: a quadratic one
ensuring closure of the gauge algebra and a linear constraint  imposed by
supersymmetry. We start presenting the latter.
A priori, the embedding tensor $\Theta_{\Ja\Jb,\Jc}{}^{\Jd}$ 
in seven dimensions is 
assigned to the ${\bf 10}\otimes{\bf 24}$
representation of ${\rm SL}(5)$. Decomposing the tensor product
\bea
{\bf 10}\otimes{\bf 24}
&=& {\bf 10} + {\bf 15} + {\bf \overline{40}} +{\bf 175}
\;,
\eea
supersymmetry restricts the embedding tensor to 
the representations ${\bf 15}+{\bf \overline{40}}\,$~\cite{deWit:2004nw,deWit:2005hv},
as we will explicitly see in the following.
It can thus be parametrized by a symmetric matrix $Y_{\Ja\Jb}=Y_{(\Ja\Jb)}$
and a tensor $Z^{\Ja\Jb,\Jc}=Z^{[\Ja\Jb],\Jc}$ with $Z^{[\Ja\Jb,\Jc]}=0$ as
\bea
\Theta_{\Ja\Jb,\Jc}{}^{\Jd}&=&
\delta^{\Jd}_{[\Ja}\,Y^{\phantom{\Jd}}_{\Jb]\Jc}
-2\epsilon_{\Ja\Jb\Jc\Je\Jf}\,Z^{\Je\Jf,\Jd}
\;.
\label{linear}
\eea
The gauge group generators~(\ref{X-theta}) in the ${\bf 5}$-representation
then take the form
\bea
(X_{\Ja\Jb}){}_{\Jc}{}^{\Jd} &=&
\Theta_{\Ja\Jb,\Jc}{}^{\Jd}~=~
\delta^{\Jd}_{[\Ja}\,Y^{\phantom{\Jd}}_{\Jb]\Jc} 
-2\epsilon_{\Ja\Jb\Jc\Je\Jf}\,Z^{\Je\Jf,\Jd}
\;.
\label{XP}
\eea
For the gauge group generators in the 
${\bf 10}$-representation
$(X_{\Ja\Jb}){}_{\Jc\Jd}{}^{\Je\Jf} = 
2 (X_{\Ja\Jb}){}^{\vphantom{\Jf]}}_{[\Jc}{}^{[\Je}_{\vphantom{\Jd]}}\delta_{\Jd]}^{\Jf]}$
we note the relation
\bea
(X_{\Ja\Jb}){}_{\Jc\Jd}{}^{\Je\Jf}+(X_{\Jc\Jd}){}_{\Ja\Jb}{}^{\Je\Jf}
~=~ 2\,Z_{\phantom{\Ja}}^{\Je\Jf,\Jg}\,d_{\Jg,[\Ja\Jb][\Jc\Jd]}
\;,
\label{Csym}
\eea
where we have defined the ${\rm SL}(5)$ invariant tensor 
$d_{\Jg,[\Ja\Jb][\Jc\Jd]}=\epsilon_{\Jg\Ja\Jb\Jc\Jd}$
in accordance with the general formulas of~\cite{deWit:2005hv},
see also~appendix~\ref{APP-modified}.
Furthermore, we note the identity
\bea
(X_{\Ja\Jb}){}_{\Jc}{}^{\Jd}+2d_{\Jc,[\Ja\Jb][\Je\Jf]}Z_{\phantom{\Ja}}^{\Je\Jf,\Jd}
&=&
\delta^{\Jd}_{[\Ja}\,Y^{\phantom{\Jd}}_{\Jb]\Jc}
\;.
\label{idY}
\eea

In addition to the linear representation constraint whose explicit solution
is given by~(\ref{linear}), a quadratic constraint needs to be imposed on
the embedding tensor in order to ensure closure of the gauge algebra.
This amounts to imposing invariance of the embedding tensor itself under
the action of the gauge group:
\bea
(X_{\Ja\Jb}){}_{\Jc\Jd}{}^{\Jg\Jh} \,\Theta_{\Jg\Jh,\Je}{}^{\Jf}
+(X_{\Ja\Jb}){}_{\Je}{}^{\Jg} \,\Theta_{\Jc\Jd,\Jg}{}^{\Jf}
-(X_{\Ja\Jb}){}_{\Jg}{}^{\Jf} \,\Theta_{\Jc\Jd,\Je}{}^{\Jg} &=& 0
\;.
\eea
Using the explicit parametrization of~(\ref{linear})
these equations reduce to
the conditions
\bea
Y_{\Ja\Jd}\,Z^{\Jd\Jb,\Jc} 
+2\epsilon_{\Ja\Je\Jf\Jg\Jh}\,Z^{\Je\Jf,\Jb}Z^{\Jg\Jh,\Jc}
&=& 0
\;,
\label{quadratic}
\eea
for the tensors $Y_{\Ja\Jb}$ and $Z^{\Ja\Jb,\Jc}$. In terms of 
${\rm SL}(5)$ representations these quadratic constraints 
have different irreducible parts in the 
${\bf \overline{5}}$, the ${\bf \overline{45}}$, and the ${\bf \overline{70}}$ representation.
In particular, they give rise to the relations
\bea
Z^{\Ja\Jb,\Jc}\,Y_{\Jc\Jd}&=&0\;,
\qquad
Z^{\Ja\Jb,\Jc}\,X_{\Ja\Jb} ~=~0
\;,
\label{Q2}
\eea
where in the second equation, $X_{\Ja\Jb}$ is taken in an arbitrary representation.
In fact, the second equation of~\eqref{Q2} already carries the full content
of the quadratic constraint.
Yet another (equivalent) version of writing the quadratic constraints~(\ref{quadratic}) is
\bea
\Big[
X_{\Ja\Jb},X_{\Jc\Jd}
\Big]
&=&
-(X_{\Ja\Jb}){}_{\Jc\Jd}{}^{\Je\Jf}\,X_{\Je\Jf}
\;,
\label{algebra}
\eea
for the generators $X_{\Ja\Jb}$ in an arbitrary representation.
This shows the closure of the gauge algebra.
The $(X_{\Ja\Jb}){}_{\Jc\Jd}{}^{\Je\Jf}$ encode the
structure constants of this algebra, 
although by virtue of~(\ref{Csym})
and the second equation of~(\ref{Q2})
they are antisymmetric only after contraction with the embedding tensor. 
Similarly, the Jacobi identities are satisfied only 
up to extra terms that are proportional to $Z^{MN,K}$
and thus also vanish under contraction with the embedding tensor.
We will come back to this in the next section.

Summarizing, a consistent gauging of the seven-dimensional theory is defined by an embedding tensor
$\Theta_{\Ja\Jb,\Jc}{}^{\Jd}$ satisfying a linear 
and a quadratic ${\rm SL}(5)$ representation constraint
which schematically read
\bea
  \Big({\mathbb{P}}_{\bf 10}  +{\mathbb{P}}_{{\bf 175}}\Big)
  \,\Theta\phantom{\,   \Theta }  &=&   0\,,   \nn\\ 
\Big( {\mathbb{P}}_{\bf \overline{5}} + {\mathbb{P}}_{\bf \overline{45}}
+ {\mathbb{P}}_{\bf \overline{70}}
  \Big)\,\Theta\,   \Theta &=&  0\,.  
  \label{sum}
\eea
The first of these equations can be explicitly solved
in terms of two tensors $Y_{\Ja\Jb}$ and $Z^{\Ja\Jb,\Jc}$ leading to~(\ref{linear});
the quadratic constraint then translates into the 
conditions~(\ref{quadratic}) on these tensors. 
In the rest of this paper we will demonstrate that
an embedding tensor $\Theta$ solving equations~(\ref{sum})
defines a consistent gauging in seven dimensions.

\section{Vector and tensor gauge fields}
\setcounter{equation}{0}
\label{SecVecTen}

We will for the gauged theory employ a formulation which apart from the 
vector fields $A_\mu^{\Ja\Jb}$ contains the two-form tensors $B_{\mu\nu}{}_{\Ja}$
and the three-form tensor fields $S_{\mu\nu\rho}^{\Ja}$, the latter
transforming in the $\overline {\bf 5}$ of ${\rm SL}(5)$.
The components of the embedding tensor $\Theta_{\Ja\Jb,\Jc}{}^{\Jd}$
will project onto those fields that are actually involved in the gauging.
In particular, the three-form tensors $S_{\mu\nu\rho}^{\Ja}$
appear always projected under $Y_{\Ja\Jb}$.
The combined vector and tensor gauge invariances together with a
topological coupling of the three-form tensors will ensure
that the number of physical degrees of freedom will remain independent of the
embedding tensor. The latter will only determine how the degrees of freedom
are distributed among the vector and the different tensor fields.
In particular, at $\Theta_{\Ja\Jb,\Jc}{}^{\Jd}=0$ one recovers the ungauged
theory of~\cite{Sezgin:1982gi} which is exclusively formulated in terms of vector
and two-form tensor fields.
The identities~(\ref{quadratic}) and their consequences~(\ref{Q2}), (\ref{algebra})
prove essential for consistency of this construction.

Already in the previous section we have encountered 
the fact that the ``structure constants'' 
$(X_{\Ja\Jb}){}_{\Jc\Jd}{}^{\Je\Jf}$
of the gauge algebra~(\ref{algebra}) are neither 
antisymmetric nor satisfy the Jacobi identities.
Both, antisymmetry and Jacobi identities are satisfied
only up to terms proportional to the tensor $Z^{\Ja\Jb,\Jc}$, i.e.~up to
terms that vanish upon contraction with the embedding tensor, cf.~(\ref{Q2}).
As a consequence, the nonabelian field strength
of the vector fields
\begin{align}
{\cal F}^{\Ja\Jb}_{\mua\mub}
&=
2 \partial_{[\mua} A_{\mub]}^{\Ja\Jb} 
+ g {(X_{\Jc\Jd})_{\Je\Jf}}_{\vphantom{[\mua}}^{\Ja\Jb} 
A_{[\mua}^{\Jc\Jd} A_{\vphantom{[]}\mub]}^{\vphantom{\Jd}\Je\Jf}
\;,
   \label{DefFsF}
\end{align}
does not transform covariantly under the standard nonabelian 
vector gauge transformations~$\delta A_{\mua}^{\Ja\Jb} 
= D_\mua \Lambda^{\Ja\Jb}$.\footnote{
  Covariant derivatives here and in the following
  refer to the ${\rm SL}(5)$ 
  index structure of the object they act on, i.e.
  \begin{align*}
     D_{\mua} \Lambda^{\Ja\Jb} &= \partial_{\mua} \Lambda^{\Ja\Jb}
      + g {X_{\Jc\Jd,\Je\Jf}}^{\Ja\Jb}  A_{\mua}^{\Jc\Jd}  \Lambda^{\Je\Jf}
  \;,
  \end{align*}
  etc.}
Consistency requires the introduction of the modified field strength
\begin{align}
   {\cal H}^{(2)\Ja\Jb}_{\mua\mub} &= 
   {\cal F}^{\Ja\Jb}_{\mua\mub} + g Z^{\Ja\Jb,\Jc} B_{\mua\mub\Jc} 
\;,
\label{H2}
\end{align}
where the gauge transformation of the two-forms~$B_{\mua\mub\Ja}$
will be chosen such (cf.~(\ref{gauge}) below)
that ${\cal H}^{(2)\Ja\Jb}_{\mua\mub}$ transforms
covariantly
\begin{align}
   \delta\, {\cal H}^{(2)\Ja\Jb}_{\mua\mub} &= 
   - g \Lambda^{\Jc\Jd} {(X_{\Jc\Jd})_{\Je\Jf}}^{\Ja\Jb}\, {\cal H}^{(2)\Je\Jf}_{\mua\mub} 
   \;.
   \label{trafH2}
\end{align}
Similarly, the fields strength of the two-form tensor fields $B_{\mua\mub\Ja}$
is modified 
by a term proportional to the three-form tensor fields~\cite{deWit:2005hv}
\begin{align}
   {\cal H}^{(3)}_{\mua\mub\muc\,\Ja} &= 3 D_{[\mua} B_{\mub\muc]\Ja} 
                                        + 6 \epsilon_{\Ja\Jb\Jc\Jd\Je} A^{\Jb\Jc}_{[\mua} 
\Big( \partial^{\vphantom{\Jd}}_{\vphantom{[]}\mub} A^{\Jd\Je}_{\muc]} 
+ \frac 2 3 g {X_{\Jf\Jg,\Jh}}^\Jd A^{\Je\Jh}_\mub A^{\Jf\Jg}_{\muc]} \Big)
+ g Y_{\Ja\Jb} S^\Jb_{\mua\mub\muc}
\;.
\label{H3}
\end{align}
As $g\rightarrow0$ one recovers from this expression the abelian field strength
$3 \partial_{[\mua} B_{\mub\muc]\Ja}+ 6 \epsilon_{\Ja\Jb\Jc\Jd\Je} A^{\Jb\Jc}_{[\mua} 
\partial^{\vphantom{\Jd}}_{\vphantom{[]}\mub} A^{\Jd\Je}_{\muc]}$
of the ungauged theory~\cite{Sezgin:1982gi}.
Again, gauge transformations of the three-forms $S^\Jb_{\mua\mub\muc}$
will be chosen such (cf.~(\ref{gauge}) below) that ${\cal H}^{(3)}_{\mua\mub\muc\,\Ja}$
transforms covariantly
\begin{align}
   \delta\, {\cal H}^{(3)}_{\mua\mub\muc\,\Ja} &= 
   g \Lambda^{\Jb\Jc} {(X_{\Jb\Jc})_{\Ja}}^\Jd \,{\cal H}^{(3)}_{\mua\mub\muc\,\Jd}
   \;.
   \label{trafH3}
\end{align}
To determine the proper transformation behavior of the tensor fields,
we first note the variation of the modified field strengths~(\ref{H2}), (\ref{H3})
under arbitrary variations of the vector and tensor fields:
\begin{align}
   \delta {\cal H}^{(2)\Ja\Jb}_{\mua\mub} &= 
   2 D^{\phantom{\Ja}}_{[\mua} (\Delta A_{\mub]}^{\Ja\Jb} )
                                             + g Z^{\Ja\Jb,\Jc} \Delta B_{\mua\mub\,\Jc} \;,
   \nonumber\\[1ex]
   \delta {\cal H}^{(3)}_{\mua\mub\muc\,\Ja} &= 
   3 D_{[\mua} (\Delta B_{\mub\muc]\Ja}) 
+ 6 \epsilon_{\Ja\Jb\Jc\Jd\Je} {\cal H}_{[\mua\mub}^{(2)\Jb\Jc} \Delta A_{\muc]}^{\Jd\Je}
+ g Y_{\Ja\Jb} \Delta S_{\mua\mub\muc}^\Jb             
\;,
   \label{VarHH}
\end{align}
where here and in the following it proves useful to define the 
``covariant variations"
\begin{align}
  \Delta A_{\mua}^{\Ja\Jb} &\equiv \delta A_{\mua}^{\Ja\Jb} 
  \;,
  \nonumber\\[1ex]
  \Delta B_{\mua\mub\,\Ja} &\equiv \delta B_{\mua\mub\,\Ja} 
  - 2 \epsilon_{\Ja\Jb\Jc\Jd\Je} A_{[\mua}^{\Jb\Jc} \,\delta A_{\mub]}^{\Jd\Je} 
  \;,
  \nonumber\\[1ex]
Y_{\Ja\Jb}\,
  \Delta S_{\mua\mub\muc}^\Jb &\equiv 
  Y_{\Ja\Jb}\,\Big(
  \delta S_{\mua\mub\muc}^\Jb
- 3 B^{\phantom{\Jb}}_{[\mua\mub\,\Jc}\, \delta A_{\muc]}^{\Jc\Jb} 
+ 2 \epsilon_{\Jc\Jd\Je\Jf\Jg} A_{[\mua}^{\Jb\Jc} A_{\mub}^{\Jd\Je} \,\delta A_{\muc]}^{\Jf\Jg}
\Big)
\;.
\label{covV}
\end{align}
In terms of these, we can now present the full set of vector and tensor 
gauge transformations:
\begin{align}
   \Delta A_{\mua}^{\Ja\Jb} &= D_\mua \Lambda^{\Ja\Jb} 
   - g Z^{\Ja\Jb,\Jc} \Xi_{\mua\,\Jc} 
   \;,
   \nonumber\\[1ex]
   \Delta B_{\mua\mub\,\Ja} &= 2 D_{[\mua} \Xi_{\mub]\,\Ja} 
             - 2 \epsilon_{\Ja\Jb\Jc\Jd\Je}\, {\cal H}_{\mua\mub}^{(2)\Jb\Jc} \Lambda^{\Jd\Je} 
             - g Y_{\Ja\Jb} \Phi_{\mua\mub}^\Jb 
   \;,
   \nonumber\\[1ex]
   Y_{\Ja\Jb}\,\,\Delta S_{\mua\mub\muc}^\Jb &= 
   Y_{\Ja\Jb}\,\Big(
   3 D^{\phantom{\Jb}}_{[\mua} \Phi_{\mub\muc]}^\Jb
                   - 3 {\cal H}_{[\mua\mub}^{(2)\Jb\Jc} \Xi_{\muc]\,\Jc} 
                   + {\cal H}^{(3)}_{\mua\mub\muc\,\Jc} \Lambda^{\Jc\Jb} 
\Big)\;,
\label{gauge}
\end{align}
with gauge parameters $\Lambda^{\Ja\Jb}$, $\Xi_{\mua\,\Ja}$,
and  $\Phi_{\mua\mub}^\Ja$, corresponding to
vector and tensor gauge transformations, respectively.
Indeed, one verifies with \eqref{VarHH}, \eqref{covV} that the 
transformations~\eqref{gauge} induce the proper
covariant transformation behavior of the modified fields 
strengths~(\ref{trafH2}), (\ref{trafH3}). 
The quadratic identities~(\ref{Q2}) play a crucial role in this 
derivation.\footnote{The gauge transformations~(\ref{gauge})
differ from the general formulas derived in~\cite{deWit:2005hv}
by a redefinition of the tensor gauge parameters
which strongly simplifies the expressions and in particular allows
to cast them into the covariant form~(\ref{covV}), (\ref{gauge}).
We give the explicit translation in appendix~\ref{APP-modified}.}
As $g\rightarrow0$ one recovers from~(\ref{gauge}) the vector and tensor gauge
transformations of the ungauged theory~\cite{Sezgin:1982gi}.
Switching on the gauging induces a covariantization $\partial\rightarrow D$,
$2\partial A\rightarrow{\cal H}$, etc.~of the formulas
together with the shifts $g Z^{\Ja\Jb,\Jc} \Xi_{\mua\,\Jc}$
and $g Y_{\Ja\Jb} \Phi_{\mua\mub}^\Jb$ in the transformation 
laws of the vector and tensor fields, respectively.
It is a nontrivial check of consistency that all the unwanted
terms in the variation can precisely be absorbed by such shifts
proportional to the components of the embedding tensor~(\ref{linear}).
This action of the tensor gauge transformations eventually allows to 
eliminate some of the vector and
tensor gauge fields by fixing part of the gauge symmetry. 
We will discuss this in more detail in section~\ref{sec:gfixing}.

It is straightforward  to verify that the gauge transformations~(\ref{gauge})
consistently close into the algebra
\begin{align}
   [ \delta_{\Lambda_1} , \delta_{\Lambda_2} ] &= 
   \delta_{\tilde{\Lambda}}+\delta_{\tilde{\Xi}}+\delta_{\tilde{\Phi}}
   \nonumber\\[1ex]
  [ \delta_{\Xi_1} , \delta_{\Xi_2} ] &= \delta_{{\Phi}}
   \;,
\end{align}
with
\begin{align}
   \tilde\Lambda^{\Ja\Jb} &= 
   g {X_{\Jc\Jd,\Je\Jf}}^{\Ja\Jb} 
   \Lambda_{[1}^{\Jc\Jd}  \Lambda_{2]}^{\vphantom{\Jd}\Je\Jf} \;,\nonumber\\[1ex]
   \tilde\Xi_{\mua\,\Ja} &= \epsilon_{\Ja\Jb\Jc\Jd\Je} \,
   \Big( \Lambda_1^{\Jb\Jc} D_\mua \Lambda_2^{\Jd\Je} 
   - \Lambda_2^{\Jb\Jc} D_\mua \Lambda_1^{\Jd\Je} \Big)     
                                                        \;,\nonumber\\[1ex]
   \tilde\Phi_{\mua\mub}^{\,\Ja} &= 
   - 2\epsilon_{\Jb\Jc\Jd\Je\Jf}\, {\cal H}_{\mua\mub}^{(2)\Je\Jf} 
                  \Lambda_{[1}^{\vphantom{\Jd}\Ja\Jb} \Lambda_{2]}^{\Jc\Jd} 
\;,\nonumber\\[1ex]
\Phi_{\mua\mub}^{\,\Ja} &= 
 g Z^{\Ja(\Jb,\Jc)} \Big(
\Xi_{1\,\mua\Jb} \Xi_{2\,\mub\Jc}-\Xi_{2\,\mua\Jb} \Xi_{1\,\mub\Jc}\Big)
\;,
\nonumber
\end{align}
with all other transformations commuting.

Let us further note that the modified field strengths~(\ref{H2}), (\ref{H3})
satisfy the following deformed Bianchi identities:
\begin{align}
   D^{\phantom{\Ja}}_{[\mua} {\cal H}_{\mub\muc]}^{(2)\Ja\Jb} &= 
   \frac13 g Z^{\Ja\Jb,\Jc} {\cal H}^{(3)}_{\mua\mub\muc\,\Jc} 
   \;,\nonumber\\[1ex]
   D^{\phantom{\Ja}}_{[\mua} {\cal H}^{(3)}_{\mub\muc\mud]\,\Ja} 
        &= \frac 3 2 \epsilon_{\Ja\Jb\Jc\Jd\Je}\, 
        {\cal H}_{[\mua\mub}^{(2)\Jb\Jc} {\cal H}_{\muc\mud]}^{(2)\Jd\Je}
           + \frac 14g Y_{\Ja\Jb} \, {\cal H}^{(4)\,\Jb}_{\mua\mub\muc\mud}
           \;,
           \label{Bianchi}
\end{align}
which will be important in the following. 
The last term on the r.h.s.\ of 
$D^{\phantom{\Ja}}_{[\mua} {\cal H}^{(3)}_{\mub\muc\mud]\,\Ja}$
is the covariant field strength
of the three-forms, defined as
\begin{align}
Y_{\Ja\Jb}\, {\cal H}^{(4)\,\Jb}_{\mua\mub\muc\mud} &= 
Y_{\Ja\Jb}  \Big( 4D^{\phantom{\Ja}}_{[\mua} S_{\mub\muc\mud]}^\Jb
+ 6 {\cal F}_{[\mua\mub}^{\Jb\Jc} B^{\phantom{\Ja}}_{\muc\mud]\Jc}  
+ 3 g Z^{\Jb\Jc,\Jd} B_{[\mua\mub\,\Jc} B_{\muc\mud]\,\Jd}
\nonumber \\ 
& \qquad\quad
               + 8 \epsilon_{\Jc\Jd\Je\Jf\Jg} A_{[\mua}^{\Jb\Jc} 
               A_{\mub}^{\Jd\Je} \partial^{\vphantom{\Ja}}_{\muc} A_{\mud]}^{\Jf\Jg}
  +4g  \epsilon_{\Jc\Jd\Je\Ji\Jj} {X_{\Jf\Jg,\Jh}}^{\Ji} 
  A_{[\mua}^{\Jb\Jc} A_\mub^{\Jd\Je} A_\muc^{\Jf\Jg} A_{\mud]}^{\Jh\Jj}
\Big)                
\;,
\label{H4}
\end{align}
such that it transforms covariantly under
gauge transformations. As the three-form tensors will appear only under the
projection with $Y_{\Ja\Jb}$, it is sufficient to only define their field strength
under that same projection.

A natural question concerns the possible gauge invariant couplings
of the vector and tensor fields in the Lagrangian. Due to the
covariant transformations~(\ref{trafH2}),~(\ref{trafH3}) 
it is obvious that gauge invariant couplings can be obtained by
properly contracting the modified field strengths~(\ref{H2}),~(\ref{H3}).
E.g.~the gauge invariant kinetic term for the two-form tensor 
fields is given by
\begin{align}
   {\cal L} \propto&~
   {\cal M}^{\Ja\Jb}\,
   {\cal H}^{(3)}_{\mua\mub\muc\,\Ja}\,{\cal H}^{(3)\mua\mub\muc}{}_{\,\Jb}
   \;,
   \label{LkinH3}
\end{align}
where the metric ${\cal M}^{\Ja\Jb}$ 
is a function of the scalar fields 
(explicitly defined in \eqref{DefCalM} below)
that transforms covariantly under gauge transformations
\begin{align}
   \delta\, {\cal M}^{\Ja\Jb} &= 
   - 2g \Lambda^{\Jc\Jd} {(X_{\Jc\Jd})_{\Je}}{}^{(\Ja}\, {\cal M}_{\vphantom{.}}^{\Jb)\Je}
   \;.
   \label{trafM}
\end{align}
Similarly, the gauge invariant kinetic term for the vector fields is given as
\begin{align}
   {\cal L} \propto&~
   {\cal M}_{\Ja\Jc} \, {\cal M}_{\Jb\Jd} \,
   {\cal H}^{(2)\Ja\Jb}_{\mua\mub}\,{\cal H}^{(2)\mua\mub\,\Jc\Jd}
   \;,
\end{align}
with ${\cal M}_{\Ja\Jc}{\cal M}^{\Jc\Jb}=\delta_{\Ja}^{\Jb}$.
In addition to these terms, there is a single unique topological
term in seven dimensions that combines vector and tensor fields in such a way
that it is invariant under the full set of nonabelian
vector and tensor gauge transformations \eqref{gauge}
up to total derivatives. It is given by
\begin{align}
   {\cal L}_{\text{VT}} =& - \frac 1 {9}\, 
   \epsilon^{\mua\mub\muc\mud\mue\muf\mug} \nonumber \\ &
     \Bigg[ g Y_{\Ja\Jb} S^\Ja_{\mua\mub\muc} \Big( D^{\vphantom{\Ja}}_\mud S_{\mue\muf\mug}^\Jb
+ \frac 3 2 g Z^{\Jb\Jc,\Jd} B_{\mud\mue\,\Jc} B_{\muf\mug\,\Jd}
+ 3 {\cal F}^{\Jb\Jc}_{\mud\mue} B^{\phantom{\Ja}}_{\muf\mug\,\Jc}
\nonumber \\ & \qquad \qquad
                 + 4 \epsilon_{\Jc\Jd\Je\Jf\Jg} A^{\Jb\Jc}_{\mud} A^{\Jd\Je}_{\mue} \partial^{\vphantom{\Ja}}_{\muf} A_\mug^{\Jf\Jg}
        + g \epsilon_{\Jc\Jd\Je\Jj\Jk} {X_{\Jf\Jg,\Jh\Ji}}^{\Jj\Jk} A_\mud^{\Jb\Jc} A_\mue^{\Jd\Je} A_\muf^{\Jf\Jg} A_\mug^{\Jh\Ji} \Big) 
  \nonumber \\  &  \quad
      + 3 g Z^{\Ja\Jb,\Jc} (D_\mua B_{\mub\muc\,\Ja}) B_{\mud\mue\,\Jb} B_{\muf\mug\,\Jc} 
    - \frac 9 2 {\cal F}_{\mua\mub}^{\Ja\Jb} B_{\muc\mud\,\Ja} D_{\mue} B_{\muf\mug\,\Jb}
  \nonumber \\ & \quad
    + 18 \epsilon_{\Ja\Jb\Jc\Jd\Je} {\cal F}_{\mua\mub}^{\Ja\Ji} A^{\Jb\Jc}_{\muc} 
            \Big( \partial^{\vphantom{\Ja}}_\mud A^{\Jd\Je}_{\mue} 
                                       + \frac 2 3 g {X_{\Jf\Jg,\Jh}}^\Jd A^{\Je\Jh}_\mud A^{\Jf\Jg}_{\mue} \Big)\, B_{\muf \mug\,\Ji}
  \nonumber \\ & \quad
    + 9 g \epsilon_{\Ja\Jb\Jc\Jd\Je} Z^{\Ja\Ji,\Jj} A^{\Jb\Jc}_{\mua} \Big( \partial^{\vphantom{\Ja}}_\mub A^{\Jd\Je}_{\muc} 
          + \frac 2 3 g {X_{\Jf\Jg,\Jh}}^\Jd A^{\Je\Jh}_\mub A^{\Jf\Jg}_{\muc} \Big)\, B_{\mud\mue\,\Ji} B_{\muf\mug\,\Jj} \nonumber \\
  & + \frac {36} 5 \epsilon_{\Ja\Jc\Jd\Jg\Jh} \; \epsilon_{\Jb\Je\Jf\Ji\Jj} 
       A_\mua^{\Ja\Jb} A_\mub^{\Jc\Jd} A_\muc^{\Je\Jf} 
             (\partial_\mud^{\vphantom{\Ja}} A_\mue^{\Jg\Jh}) (\partial_\muf^{\vphantom{\Ja}} A_\mug^{\Ji\Jj})
     \nonumber \\ &   
    + 8 g \epsilon_{\Ja\Jc\Jd\Je\Jf} \; \epsilon_{\Jb\Jg\Jh\Jm\Jn} {X_{\Ji\Jj,\Jk\Jl}}^{\Jm\Jn}
          A_\mua^{\Ja\Jb} A_\mub^{\Jc\Jd} A_\muc^{\Jg\Jh} A_\mud^{\Ji\Jj} A_\mue^{\Jk\Jl} 
\partial_\muf^{\vphantom{\Ja}} A_\mug^{\Je\Jf} \nonumber \\
  & - \frac 4 7 g^2 \epsilon_{\Ja\Jc\Jd\Jo\Jp} \; \epsilon_{\Jb\Ji\Jj\Jq\Jr} {X_{\Je\Jf,\Jg\Jh}}^{\Jo\Jp} {X_{\Jk\Jl,\Jm\Jn}}^{\Jq\Jr}
        A_\mua^{\Ja\Jb} A_\mub^{\Jc\Jd} A_\muc^{\Je\Jf} A_\mud^{\Jg\Jh} A_\mue^{\Ji\Jj} A_\muf^{\Jk\Jl} A_\mug^{\Jm\Jn}
\Bigg]
\;.
\label{VT}
\end{align}
As $g\rightarrow0$ this topological term reduces to the ${\rm SL}(5)$ invariant
Chern-Simons term of the ungauged theory~\cite{Sezgin:1982gi}.
Upon switching on the gauging, gauge invariance under the 
extended nonabelian transformations~\eqref{gauge} requires an extension of this
Chern-Simons term which in particular
includes a first order kinetic term for the three-form tensor 
fields~$S^\Ja_{\mua\mub\muc}$.
Again these tensor fields appear only under projection
with the tensor $Y_{\Ja\Jb}$.
Under variation of the vector and tensor fields, the
vector-tensor Lagrangian~(\ref{VT}) transforms as
\begin{align}
   \delta {\cal L}_{\text{VT}} = &
      - \frac 1 {18}\, \epsilon^{\mua\mub\muc\mud\mue\muf\mug} \Bigg[
         Y_{\Ja\Jb} \, {\cal H}^{(4)\Ja}_{\mua\mub\muc\mud} \, \Delta S^\Jb_{\mue\muf\mug} 
                + 6 {\cal H}^{(2)\Ja\Jb}_{\mua\mub} {\cal H}^{(3)}_{\muc\mud\mue\Ja} \,\Delta B_{\muf\mug\Jb}
\nonumber \\ & \qquad \qquad \qquad \qquad 
                - 2 {\cal H}^{(3)}_{\mua\mub\muc\Ja} {\cal H}^{(3)}_{\mud\mue\muf\Jb} \,\Delta A^{\Ja\Jb}_\mug \Bigg]
\; +{\rm total~derivatives} \;,
\label{varCS}
\end{align}
in terms of the ``covariant variations''~(\ref{covV}).
With~(\ref{gauge}) one explicitly verifies that this variation 
reduces to a total derivative. 
To show this one needs the deformed Bianchi identities \eqref{Bianchi} 
as well as the ${\rm SL}(5)$ relation
\begin{align}
     R_1^{[\Ja\Jb} R_2^{\vphantom{[]}\Jc\Jd}\, R_3^{\Je]\Jf} + 
     R_2^{[\Ja\Jb} R_3^{\vphantom{[]}\Jc\Jd}\, R_1^{\Je]\Jf} + 
     R_3^{[\Ja\Jb} R_1^{\vphantom{[]}\Jc\Jd}\, R_2^{\Je]\Jf}
      &= 0  \;,
   \label{RelSL5AAA}
\end{align}
for arbitrary tensors $R^{\Ja\Jb}_{1,2,3} = R^{[\Ja\Jb]}_{1,2,3}$.
\footnote{In terms of representations, this is the 
statement that the threefold symmetric product of three ${\bf 10}$
representations of ${\rm SL}(5)$ does not contain a ${\bf 5}$.}

\section{Coset space structure and the $T$-tensor}
\setcounter{equation}{0}
\label{sec:CT}

In this section we introduce the scalar sector of maximal seven-dimensional
supergravity, which is described in terms of the scalar coset space
${\rm SL}(5)/{\rm SO}(5)$. This allows to manifestly realize the 
global ${\rm SL}(5)$ symmetry of the ungauged theory
while the local ${\rm SO}(5)\sim{\rm USp}(4)$ symmetry coincides
with the $R$-symmetry of the theory. 
For the gauged theory we further introduce the $T$-tensor 
as the ${\rm USp}(4)$ covariant analog 
of the embedding tensor $\Theta$. 
This tensor will be of importance later since its
irreducible components naturally couple to the fermions, all of which
come in representations of the $R$-symmetry group.

\subsection{The ${\rm SL}(5)/{\rm SO}(5)$ coset space}
\label{sec:coset}

The scalar fields in seven dimensions parametrize the
coset space ${\rm SL}(5)/{\rm SO}(5)$. 
They are most conveniently 
described by a matrix ${\cal V}\in {\rm SL}(5)$
which transforms according to
\begin{align}
   {\cal V} \; &\rightarrow \; G \, {\cal V} \, H(x) &
   G \in {\rm SL}(5), \quad
   H(x) \in {\rm SO}(5) \; ,
   \label{TrafoCalV}
\end{align}
under global ${\rm SL}(5)$ and local ${\rm SO}(5)$ 
transformations, respectively 
(see~\cite{deWit:2002vz} for an introduction to 
the coset space structures in supergravity theories).
The local ${\rm SO}(5)$ symmetry
reflects the coset space structure of the scalar target space,
the corresponding connection is a composite field.
One can impose a gauge condition with respect to the 
local ${\rm SO}(5)$ invariance which amounts to 
fixing a coset representative,
i.e.\ a minimal parametrization of the coset space
in terms of the $14=24-10$ physical scalars. 
This induces a nonlinear realization of the 
global ${\rm SL}(5)$ symmetry obscuring the
group theoretical structure and complicating the calculations.
It is therefore most convenient to postpone this gauge fixing 
till the end.

In particular, 
the formulation~(\ref{TrafoCalV}) is indispensable
to describe the coupling to fermions 
with the group ${\rm SO}(5)\sim{\rm USp}(4)$
acting as the $R$-symmetry group of the theory. 
For ${\rm USp}(4)$ we use indices $\ja, \, \jb, \, \ldots = 1, \ldots, 4$
to label its fundamental representation.
The ${\rm USp}(4)$ invariant symplectic form $\Omega_{\ja\jb}$ has the properties
\begin{align}
   \Omega_{\ja\jb} &= \Omega_{[\ja\jb]} &
   (\Omega_{\ja\jb})^* &= \Omega^{\ja\jb} &
   \Omega_{\ja\jb} \Omega^{\jc\jb} &= \delta_\ja^\jc \; .
\end{align}
The lowest ``bosonic" ${\rm USp}(4)$ representations are defined in terms of the
fundamental representation (\!$\tinyyoung{\cr}$\,) with index structures according to
  \begin{align}
     {\bf 1}:  & ~~\cdot  & V_{\bf 1} \nonumber\\[.5ex]
     {\bf 5}:  & ~\tinyyoung{ \cr \cr} &   {V_{\bf 5}}^{\ja\jb} &= {V_{\bf 5}}^{[\ja\jb]} \;,
                                  &  \Omega_{\ja\jb } {V_{\bf 5}}^{\ja\jb} &= 0\;, \nonumber\\[.5ex]
     {\bf 10}: & ~\tinyyoung{&  \cr }
     &  {V_{\bf 10}}^{\ja\jb} &= {V_{\bf 10}}^{(\ja\jb)}\;, \nonumber\\[.5ex]
     {\bf 14}: &~\tinyyoung{&  \cr & \cr}
     & {{V_{\bf 14}}^{\ja\jb}}_{\jc\jd} &= {{V_{\bf 14}}^{[\ja\jb]}}_{[\jc\jd]} \;,
      &
                  {{V_{\bf 14}}^{\ja\jb}}_{\jc\jb} &= 0 \;,&
\Omega_{\ja\jb} {{V_{\bf 14}}^{\ja\jb}}_{\jc\jd} &= 0 = 
\Omega^{\jc\jd} {{V_{\bf 14}}^{\ja\jb}}_{\jc\jd} \;,\nonumber\\[.5ex]
     {\bf 35}: &~
     \tinyyoung{& & \cr \cr}
& {{V_{\bf 35}}^{\ja\jb}}_{\jc\jd} &= {{V_{\bf 35}}^{[\ja\jb]}}_{(\jc\jd)}\;, &
                 {{V_{\bf 35}}^{\ja\jb}}_{\jc\jb} &= 0 \;,&
\Omega_{\ja\jb} {{V_{\bf 35}}^{\ja\jb}}_{\jc\jd} &= 0 \;.
\label{USp4Reps}
  \end{align}
All objects in these representations are pseudoreal, 
i.e.~they satisfy reality constraints
\begin{align}
   (V_{\bf 1})^* &= V_{\bf 1}\;,\quad
   ({V_{\bf 5}}^{\ja\jb})^* = \Omega_{\ja\jc} \Omega_{\jb\jd} {V_{\bf 5}}^{\jc\jd}\;,\quad
   ({{V_{\bf 14}}^{\ja\jb}}_{\jc\jd})^* = 
   \Omega_{\ja\je} \Omega_{\jb\jf} \Omega^{\jc\jg} \Omega^{\jd\jh} \,
                                                    {{V_{\bf 14}}^{\je\jf}}_{\jg\jh} 
                                                    \;,
   \label{DefReality}    
\end{align}
etc. We use complex conjugation to raise and lower ${\rm USp}(4)$ indices.
According to \eqref{DefReality} pseudoreal objects are defined such that 
their indices are equivalently raised and lowered  using $\Omega_{\ja\jb}$ and $\Omega^{\ja\jb}$.

Under its subalgebra $\mathfrak{usp}(4)$ the algebra $\mathfrak{sl}(5)$ splits 
as ${\bf 24}\rightarrow {\bf 10}+{\bf 14}$
into its compact and noncompact part, respectively. 
The elements $L=L_\Ja{}^\Jb t^{\Ja}{}_{\Jb}$ accordingly decompose as
\begin{align}
   {L_{\ja\jb}}^{\jc\jd} &~=~  
   2 \Lambda^{\vphantom{[\jc]}}_{[\ja}{}_{\vphantom{[\jc]}}^{[\jc} \delta^{\jd]}_{\jb]} 
   + {\Sigma^{\jc\jd}}_{\ja\jb}\;.
   \label{AlgebraSplit1}
\end{align}
The ${\rm SL}(5)$ vector indices $M$ are now represented as antisymmetric,
symplectic traceless index pairs $[\ja\jb]$ of ${\rm USp}(4)$.
In accordance with (\ref{USp4Reps}), $\Lambda$ and $\Sigma$ satisfy
$\Lambda_{[\ja}{}^{\jc}\,\Omega_{\jb]\jc}=0$, $\Sigma^{\ja\jb}{}_{\jc\jb}=0$,
${\Sigma^{\ja\jb}}_{\jc\jd} \, \Omega^{\jc\jd} = 0=\Omega_{\ja\jb} \, {\Sigma^{\ja\jb}}_{\jc\jd}$.
Note that this in particular implies the relation
\begin{align}
   \Omega_{\ja\je}\Omega_{\jb\jf}\Sigma^{\je\jf}{}_{\jc\jd} &= 
   \Omega_{\jc\je}\Omega_{\jd\jf}\Sigma^{\je\jf}{}_{\ja\jb} \; ,
\end{align}
i.e.\ viewed as a $5 \times 5$ matrix $\Sigma$ is symmetric.
In the split~(\ref{AlgebraSplit1}), the commutator (\ref{sl5}) 
between two elements $L_{1}=(\Lambda_{1},\Sigma_{1})$, $L_{2}=(\Lambda_{2},\Sigma_{2})$
takes the form
\begin{align}
   [L_1,L_2] &= L\;,
\end{align}
with $L=(\Lambda,\Sigma)$ according to
\begin{align}
{\Lambda_{\ja}}^\jb =&~ 
{{\Sigma_{1}}^{\jd\je}}_{\ja\jc} \, {{\Sigma_{2}}^{\jb\jc}}_{\jd\je} 
-{{\Sigma_{2}}^{\jd\je}}_{\ja\jc} \, {{\Sigma_{1}}^{\jb\jc}}_{\jd\je} 
+ {\Lambda_{1\;\ja}}^\jc \, {\Lambda_{2\;\jc}}^\jb - {\Lambda_{2\;\ja}}^\jc \, {\Lambda_{1\;\jc}}^\jb
\;,
   \nonumber \\[1ex]
   {{\Sigma}^{\jc\jd}}_{\ja\jb} =&~ 
                             - 2 {{{\Sigma_{1}}^{\je[\jc}}_{\ja\jb}} \, {\Lambda_{2\;\je}}^{\jd]} 
                             + 2 {{{\Sigma_{1}}^{\jc\jd}}_{\je[\ja}} \, {\Lambda_{2\;\jb]}}^\je
                             + 2 {{{\Sigma_{2}}^{\je[\jc}}_{\ja\jb}} \, {\Lambda_{1\;\je}}^{\jd]} 
                             - 2 {{{\Sigma_{2}}^{\jc\jd}}_{\je[\ja}} \, {\Lambda_{1\;\jb]}}^\je
                             \;.
   \label{SplitCom}    
\end{align}

The scalars of the supergravity multiplet parametrize the coset space ${\rm SL}(5)/{\rm SO}(5)$.
They are described by an ${\rm SL}(5)$ valued matrix 
${\cal V}_{\Ja}{}^{\ja\jb}={\cal V}_{\Ja}{}^{[\ja\jb]}$ with 
${\cal V}_{M}{}^{ab}\,\Omega_{ab}=0$.
Infinitesimally, the transformations~(\ref{TrafoCalV}) take the form
\begin{align}
   \delta {{\cal V}_{\Ja}}^{\ja\jb} &= {L_\Ja}^\Jb {{\cal V}_\Jb}^{\ja\jb} 
   +2 {{\cal V}_\Ja}^{\jc[\ja} {\Lambda_\jc}^{\jb]}(x) \;,&
    L \in \mathfrak{sl}(5) \;,\quad  \Lambda(x) \in  \mathfrak{usp}(4) \;.
    \label{TrafoCalV2}
\end{align}
The gauged theory is formally invariant under ${\rm SL}(5)$ transformations
only if the embedding tensor (\ref{X-theta}) is treated as a
spurionic object that simultaneously transforms under ${\rm SL}(5)$. Once $\Theta$ is frozen
to a constant, the theory remains invariant under local 
${\rm G}_{0}\times{\rm USp}(4)$ transformations
\begin{align}
   \delta {{\cal V}_{\Ja}}^{\ja\jb} &= 
   g\Lambda^{\Jc\Jd}(x)\,X_{\Jc\Jd,\Ja}{}^{\Jb}\, {{\cal V}_\Jb}^{\ja\jb} 
   +2 {{\cal V}_\Ja}^{\jc[\ja} {\Lambda_\jc}^{\jb]}(x)  \;,
    \label{TrafoCalV3}
\end{align}
parametrized by matrices $\Lambda^{\Ja\Jb}(x)$ and ${\Lambda_\ja}^{\jb}(x)$, respectively.

The inverse of ${\cal V}_{\Ja}{}^{\ja\jb}$ is denoted by ${{\cal V}_{\ja\jb}}^\Ja$, i.e.
\begin{align}
   {{\cal V}_\Ja}^{\ja\jb}  {{\cal V}_{\ja\jb}}^\Jb &= \delta_\Ja^\Jb\;, &
   {{\cal V}_{\ja\jb}}^\Ja  {{\cal V}_\Ja}^{\jc\jd} 
        &= \delta_{\ja\jb}^{\jc\jd}  - \frac 1 4 \Omega_{\ja\jb} \Omega^{\jc\jd} \; .
\end{align}
Later on we need to consider the variation of ${\cal V}$, 
for example in order to derive field equations from the Lagrangian or to minimize the
scalar potential.
Since ${\cal V}$ is a group element, 
an arbitrary variation can be expressed as a right multiplication 
with an algebra element of ${\rm SL}(5)$
\begin{align*}
   \delta {{\cal V}_\Ja}^{\ja\jb} &= {{\cal V}_\Ja}^{\jc\jd} {L_{\jc\jd}}^{\ja\jb}(x)
   ~=~ {{\cal V}_\Ja}^{\jc\jd}\,{\Sigma^{\ja\jb}}_{\jc\jd}(x)
   - 2{{\cal V}_\Ja}^{\jc[\ja}\,
    \Lambda_{\jc}{}^{\jb]}(x) 
\; .
\end{align*}
Since the last term simply describes a
${\rm USp}(4)$ gauge transformation which 
leaves the Lagrangian invariant
it will be sufficient to consider general variations of the type
\begin{align}
   \delta_{\Sigma} {{\cal V}_\Ja}^{\ja\jb} &= 
    {{\cal V}_\Ja}^{\jc\jd}\,{\Sigma^{\ja\jb}}_{\jc\jd}(x)
   \; .
   \label{SigmaVariations}
\end{align}
The $14$ parameters of $\Sigma$ correspond to variation
along the manifold ${\rm SL}(5)/{\rm SO}(5)$.

Finally, we introduce the scalar currents $P_\mua$ and $Q_\mua$
that describe the gauge covariant space-time derivative of the scalar fields.
Taking values in the Lie algebra $\mathfrak{sl}(5)$ they are defined as
\begin{align}
   {{\cal V}_{\ja\jb}}^\Ja \left( \partial_\mua {{\cal V}_{\Ja}}^{\jc\jd} 
                             - g A_\mua^{\Jc\Jd} {X_{\Jc\Jd,\Ja}}^\Jb {{\cal V}_\Jb}^{\jc\jd} \right)
             &~\equiv~ {P_{\mua\,\ja\jb}}^{\jc\jd} 
             + 2 {Q^{\vphantom{\jd]}}_{\mua\,[\ja}}{}_{\vphantom{\jb]}}^{[\jc} \delta^{\jd]}_{\jb]} 
             \;,
   \label{DefPQ}      
\end{align}
in accordance with the split \eqref{AlgebraSplit1}.
The transformation behavior of these currents is derived directly
from~\eqref{TrafoCalV3} and shows that they
are invariant under local ${\rm G}_{0}$ transformations.
Under local ${\rm USp}(4)$ transformations \eqref{TrafoCalV2},
${P_{\mua\,\ja\jb}}^{\jc\jd}$ transforms in the ${\bf 14}$, 
while ${Q_{\mua\,\ja}}^{\jb}$
transforms like a ${\rm USp}(4)$ gauge connection 
\begin{align}
   \delta {Q_{\mua\,\ja}}^\jb &= D_\mua {\Lambda_\ja}^\jb 
                             = \nabla_\mua {\Lambda_\ja}^\jb + {Q_{\mua\,\ja}}^\jc {\Lambda_\jc}^\ja
- {Q_{\mua\,\jc}}^\jb {\Lambda_\ja}^\jc
\;.
\end{align}
Thus $Q_\mua$ takes the role of a composite 
gauge field for the local ${\rm USp}(4)$ symmetry
and as such it appears in the covariant derivatives of all objects that transform under ${\rm USp}(4)$,
for example
\begin{align}
   D_\mua \psi^\ja &~=~ \nabla_\mua \psi^\ja - {Q_{\mua\,\jb}}^\ja \psi^\jb \nonumber \\[.5ex]
   D_\mua {P_{\mub\,\ja\jb}}^{\jc\jd}  &~=~ \nabla_\mua {P_{\mub\,\ja\jb}}^{\jc\jd}
                                        + 2 {Q_{\mua\,\je}}^{[\jc} {P_{\mub\,\ja\jb}}^{\jd]\je}
- 2 {Q_{\mua\,[\ja}}^\je {P_{\mub\jb]\je}}^{\jc\jd}  \nonumber \\[.5ex]
   D_\mua {{\cal V}_{\Ja}}^{\ja\jb} &~=~ \nabla_\mua {{\cal V}_{\Ja}}^{\ja\jb} 
+ 2 {Q_{\mua\,\jc}}^{[\ja} {\cal V}_\Ja^{\jb]\jc} 
- g A_\mua^{\Jc\Jd} {X_{\Jc\Jd\Ja}}^\Jb {{\cal V}_\Jb}^{\ja\jb}
~=~{{\cal V}_{\Ja}}^{\jc\jd}\,{P_{\mua\,\jc\jd}}^{\ja\jb}\;,
\end{align}
where $\psi^\ja$ is an arbitrary object in the fundamental representation of ${\rm USp}(4)$.

\subsection{The $T$-tensor}

All bosonic fields of the theory come in representations of 
${\rm SL}(5)$ while all fermionic fields come in representations of ${\rm USp}(4)$.
The object mediating between them is the scalar matrix ${\cal V}_{\Ja}{}^{\ja\jb}$.
E.g.\ it is convenient to define the ${\rm USp}(4)$ covariant
field strengths
\begin{align}
   {\cal H}^{(2)\ja\jb}_{\mua\mub} &\equiv \sqrt{2} \, \Omega_{\jc\jd} \, 
   {\cal V}_{\Ja}{}^{\ja\jc} {\cal V}_{\Jb}{}^{\jb\jd} \,
                                              {\cal H}^{(2)\Ja\Jb}_{\mua\mub} \;, &
   {\cal H}^{(3)}_{\mua\mub\muc\,\ja\jb} &\equiv  {{\cal V}_{\ja\jb}}^\Ja \, 
   {\cal H}^{(3)}_{\mua\mub\muc\,\Ja} \, ,
\end{align}
which naturally couple to the fermion fields.
More generally, the scalar matrix ${\cal V}_{\Ja}{}^{\ja\jb}$
maps tensors $R_{\Ja}$ and $S^{\Ja}$ in the 
${\rm SL}(5)$ representations ${\bf 5}$ and ${\bf\overline{5}}$,
respectively, into (scalar field dependent) tensors 
$R_{[\ja\jb]}$, $S^{[\ja\jb]}$ in the ${\bf 5}$ of ${\rm USp}(4)$ as
\begin{align}
   R_{[\ja\jb]} &=  {\cal V}_{\ja\jb}{}^{\Ja} R_{\Ja}\;,
   \qquad
      S^{[\ja\jb]} =  {\cal V}_{\Ja}{}^{\ja\jb} \,S^{\Ja}\;.
\end{align}
Similarly, tensors $R_{\Ja\Jb}$, $S^{\Ja\Jb}$ in the 
${\rm SL}(5)$ representations  ${\bf 10}$
and ${\bf \overline{10}}$, respectively, give rise to 
(scalar field dependent) tensors $R_{(\ja\jb)}$,
$S^{(\ja\jb)}$
in the ${\bf 10}$ of ${\rm USp}(4)$ as follows
\begin{align}
   R_{\ja\jb} &= \sqrt{2} \, \Omega^{\jc\jd} \, 
   {\cal V}_{\ja\jc}{}^{\Ja} {\cal V}_{\jb\jd}{}^{\Jb} \, R_{\Ja\Jb} &
   \Leftrightarrow &&
   R_{\Ja\Jb} &= - \sqrt{2} \, {\cal V}_{\Ja}{}^{\ja\jb} {\cal V}_{\Jb}{}^{\jc\jd} \, 
                            \delta^\je_{[\ja} \Omega^{\phantom{\ja}}_{\jb][\jc} \delta^\jf_{\jd]} \, R_{\je\jf} 
\;,
\nonumber \\[1ex]
   S^{\ja\jb} &= \sqrt{2} \Omega_{\jc\jd} 
   {\cal V}_{\Ja}{}^{\ja\jc} {\cal V}_{\Jb}{}^{\jb\jd} S^{\Ja\Jb} &
   \Leftrightarrow &&
   S^{\Ja\Jb} &= - \sqrt{2} \, {\cal V}_{\ja\jb}{}^{\Ja} {\cal V}_{\jc\jd}{}^{\Jb} \, 
                            \delta_\je^{[\ja} \Omega_{\phantom{\ja}}^{\jb][\jc} \delta_\jf^{\jd]} \, S^{\je\jf} \;,
\end{align}
where the normalization is chosen such that $R_{\ja\jb} S^{\ja\jb} = R_{\Ja\Jb} S^{\Ja\Jb}$.

Applying the analogous map to the embedding tensor $\Theta_{\Ja\Jb,\Jc}{}^{\Jd}$
(\ref{linear}) leads to the $T$-tensor~\cite{deWit:2002vt}
\begin{align}
  T_{(\je\jf)\,[\ja\jb]}{}^{[\jc\jd]} &\equiv   \sqrt{2}\,
  {\cal V}^{\Ja}{}_{\je\jg}{\cal V}^{\Jb}{}_{\jf\jh}\,\Omega^{\jg\jh}\,
                                              {\cal V}^{\Jc}{}_{\ja\jb} \,
                                              \Theta_{\Ja\Jb,\Jc}{}^{\Jd}\,{\cal V}_{\Jd}{}^{\jc\jd} 
\nonumber \\[1ex]
                          & =   \sqrt{2}\,\Omega_{\phantom{(\je}}^{\jh[\jc}\,{\delta}^{\jd]}_{(\je}
\,{\cal V}^{\Ja}{}^{\phantom{\jd]}}_{\jf)\jh}\,
                                {\cal V}^{\Jb}{}_{\ja\jb}\,Y^{\phantom{\ji}}_{\Ja\Jb} 
\nonumber \\ &\qquad \qquad \qquad
                           -2\sqrt{2}\,\epsilon_{\Ja\Jb\Jc\Jd\Je}\,Z^{\Jc\Jd,\Jf}\,
                           {\cal V}^{\Ja}{}_{\je\jg}{\cal V}^{\Jb}{}_{\jf\jh}\,
                             {\cal V}^{\Je}{}_{ab}\,{\cal V}_{\Jf}{}^{cd}\,\Omega^{gh}  \; .
\label{TT}
\end{align}
We shall see in the next section, that this tensor 
encodes the fermionic mass matrices as well as the scalar potential
of the Lagrangian.
This has first been observed for the $T$-tensor in the maximal $D=4$ 
supergravity~\cite{deWit:1982ig}.

Recall that the components  $Y_{\Ja\Jb}$ and $Z^{\Ja\Jb,\Jc}$ of $\Theta$
transform in the ${\bf 15}$ and the $\overline{\bf 40}$ of ${\rm SL}(5)$, respectively.
Under ${\rm USp}(4)$ they decompose as
\begin{align}
{\bf 15}+{\bf\overline{40}} & ~\rightarrow~ ({\bf 1}+{\bf 14})+  ({\bf 5}+{\bf 35}) \;.
\end{align}
Accordingly, the $T$-tensor can be decomposed into its four ${\rm USp}(4)$ irreducible components
that we denote by $B$, ${B^{[\ja\jb]}}_{[\jc\jd]}$, $C_{[\ja\jb]}$, and ${C^{[\ja\jb]}}_{(\jc\jd)}$,
respectively, with index structures according to~(\ref{USp4Reps}). This yields
\begin{align}
  T_{(\je\jf)\,\ja\jb}{}^{\jc\jd}
  =&~
  \ft12 B\,\Omega^{\vphantom{\jb}}_{\ja(\je}\,\delta_{\jf)}^{[\jc} \delta^{\jd]}_{\jb}
  - \ft12 B\,\Omega^{\vphantom{\jb}}_{\jb(\je}\,\delta_{\jf)}^{[\jc} \delta^{\jd]}_{\ja}
  + \delta_{(\je}^{[\jc}\,\Omega^{\vphantom{\jb}}_{\jf)\jg}\,B^{\jd]\jg}{}_{\ja\jb}
  \nonumber\\[1ex]
  &
  + \ft12 \, C^{\vphantom{\jb}}_{\ja(\je}\,\delta_{\jf)}^{[\jc} \delta^{\jd]}_{\jb}
  - \ft12 \,C^{\vphantom{\jb}}_{\jb(\je}\,\delta_{\jf)}^{[\jc} \delta^{\jd]}_{\ja}
  -\ft18 \Omega^{\jc\jd}\,C^{\vphantom{\jb}}_{\ja(\je}\,\Omega_{\jf)\jb}
  +\ft18 \Omega^{\jc\jd}\,C^{\vphantom{\jb}}_{\jb(\je}\,\Omega_{\jf)\ja}
  \nonumber\\[1ex]
  &
  + \ft14\Omega_{\ja\jb}\,C^{\vphantom{\jb}}_{\jg(\je}\,\delta_{\jf)}^{[\jc}\,
  \Omega^{\jd]\jg}_{\phantom{\jf}}
  + \ft12 \Omega_{\je[\ja}\,C^{\jc\jd}{}_{\jb]\jf}
  + \ft12 \Omega_{\jf[\ja}\,C^{\jc\jd}{}_{\jb]\je}
  + \ft14 \Omega_{\ja\jb}\,C^{\jc\jd}{}_{\je\jf}
  \;.
  \label{TABCD}
\end{align}
In appendices~\ref{app:tensors}, \ref{app:TQC} we present a more systematic account to these
decompositions in terms of ${\rm USp}(4)$ projection operators
which simplify the calculations.
In particular, the parametrization~(\ref{TABCD})
takes the compact form~$\eqref{TTwithTaus}$. 

For the components $Y_{\Ja\Jb}$ and $Z^{\Ja\Jb,\Jc}$ the parametrization~(\ref{TABCD}) 
yields explicitly
\begin{align}
   Y_{\Ja\Jb} &={{\cal V}_\Ja}^{\ja\jb} {{\cal V}_\Jb}^{\jc\jd}\,Y_{\ja\jb,\jc\jd}\;,
   &\qquad&
   Z^{\Ja\Jb,\Jc} = 
   \sqrt{2} {{\cal V}_{\ja\jb}}^\Ja {{\cal V}_{\jc\jd}}^\Jb {{\cal V}_{\je\jf}}^\Jc \Omega^{\jb\jd}
                                      Z^{(\ja\jc)[\je\jf]}\;, \nonumber\\[2ex]
 &  \qquad\qquad    \text{with}   
 &Y_{\ja\jb,\jc\jd}&=  \frac 1 {\sqrt{2}}  
                 \Big[  ( \Omega_{\ja\jc} \Omega_{\jb\jd} - \ft 1 4 \Omega_{\ja\jb} \Omega_{\jc\jd} )\,B
+ \Omega_{\ja\je} \Omega_{\jb\jf} {B^{[\je\jf]}}_{[\jc\jd]} \Big]\;, \nonumber\\[.5ex]
&  & 
   Z^{(\ja\jb)[\jc\jd]}& =\frac 1 {16} \Omega^{\ja[\jc} C^{\jd]\jb} 
                              + \frac 1 {16} \Omega^{\jb[\jc} C^{\jd]\ja} 
                     - \frac 1 8 \Omega^{\ja\je} \Omega^{\jb\jf} {C^{\jc\jd}}_{\je\jf}
                     \;,
\label{YZABCD}
\end{align}
where $C^{\ja\jb} = \Omega^{\ja\jc} \Omega^{\jb\jd} C_{\jc\jd}$.
Note that $\Theta$ and thus $Y_{\Ja\Jb}$ and $Z^{\Ja\Jb,\Jc}$ are constant matrices.
In contrast, the $T$-tensor and thus the tensors $B$, $C$ are functions of the scalar fields. 
It is useful to give also the inverse relations
\begin{align}
   B &~= ~
   \frac {\sqrt{2}} 5 \Omega^{\ja\jc} \Omega^{\jb\jd}  Y_{\ja\jb,\jc\jd}
       \;,
\nonumber\\[1ex]
      {B^{\ja\jb}}_{\jc\jd} &~=~ 
      \sqrt{2}\, \Big[
    \Omega^{\ja\je} \Omega^{\jb\jf} \delta^{\jg\jh}_{\jc\jd}
            -\ft15 \left( \delta^{\ja\jb}_{\jc\jd} - \ft 1 4 \Omega^{\ja\jb} \Omega_{\jc\jd} \right)
\Omega^{\je\jg} \Omega^{\jf\jh}   \Big]  \,
Y_{\je\jf,\jg\jh}
             \;,
       \nonumber \\[1ex]
   C^{\ja\jb} &~=~ 8 \,\Omega_{\jc\jd}\,Z^{(\ja\jc)[\jb\jd]}\;, \nonumber \\[1ex]
   {C^{\ja\jb}}_{\jc\jd} 
        &~=~   8 \left( - \Omega_{\jc\je}\Omega_{\jd\jf} \delta^{\ja\jb}_{\jg\jh}  \, 
               + \, \Omega^{\phantom{\ja}}_{\jg(\jc} \delta^{\ja\jb}_{\jd)\je} \Omega_{\jf\jh}  \right)
Z^{(\je\jf)[\jg\jh]}
\;.   
\label{BC}
\end{align}
Under the variation \eqref{SigmaVariations} 
of the scalar fields, these tensors transform as
\bea
\delta_\Sigma \, B &=& -\ft25\, \Sigma^{\ja\jb}{}_{\jc\jd}\,B^{\jc\jd}{}_{\ja\jb} 
\;,
\nonumber\\[1ex]
\delta_\Sigma \, B^{\ja\jb}{}_{\jc\jd} &=&
-2\,B\,\Sigma^{\ja\jb}{}_{\jc\jd}
-
\Sigma^{\ja\jb}{}_{\jg\jh}\,B^{\jg\jh}{}_{\jc\jd}
-\Sigma^{\jg\jh}{}_{\jc\jd}\,B^{\ja\jb}{}_{\jg\jh}+
\ft25
(\delta^{\ja\jb}_{\jc\jd}-\ft14\Omega^{\ja\jb}\Omega_{\jc\jd}) 
             \Sigma^{\je\jf}{}_{\jg\jh}\,B^{\jg\jh}{}_{\je\jf}
\;,
\nonumber\\[1ex]
\delta_\Sigma \, C^{\ja\jb} &=& 
   \ft12\, \Sigma^{\ja\jb}{}_{\jc\jd}\,C^{\jc\jd} 
    +2\, \Omega^{\je[\ja} \Sigma^{\jb]\jf}{}_{\jc\jd}\,C^{\jc\jd}{}_{\je\jf} 
\;,
\nonumber\\[1ex]
\delta_\Sigma \, C^{\ja\jb}{}_{\jc\jd} &=&
    4\,\Omega^{\jg[\ja}\,\Sigma^{\jb]\jh}{}_{\jg(\jc}\, C_{\jd)\jh}
   + \Omega^{\jg[\ja}_{\phantom{(\jc}}\,\delta^{\jb]}_{(\jc}\,\Sigma^{\jk\jh}{}_{\jd)\jg}\, C_{\jk\jh}
    +\Omega^{\jg\jk}\,\delta^{[\ja}_{(\jc}\,\Sigma^{\jb]\jh}{}_{\jd)\jg}\, C_{\jk\jh}
\nonumber\\[.5ex]
&&{}
+
\Sigma^{\ja\jb}{}_{\jg\jh}\,C^{\jg\jh}{}_{\jc\jd}
+\Sigma^{\jk[\ja}{}_{\jg\jh}\,\delta^{\jb]}_{(\jc}\,C^{\jg\jh}{}_{\jd)\jk}
\nonumber\\[.5ex]
&&
{}
+4\,\Sigma^{\jk\jm}{}_{\jl(\jc}\,\Omega_{\jd)\jk}\,\Omega^{\jn[\ja}\,C^{\jb]\jl}{}_{\jm\jn}
- \delta^{[\ja}_{(\jc}\,\Omega^{\phantom{\ja}}_{\jd)\jk}\,\Omega^{\jb]\jn}\,
\Sigma^{\jk\jm}{}_{\jl\jg}\,C^{\jg\jl}{}_{\jm\jn}
\;.
\label{varABCD}
\eea
These variations will be relevant in the next section, 
since  in the Lagrangian the tensors $B$, $C$ appear
in the fermionic mass matrices and in the scalar potential.
Furthermore, one derives from~\eqref{varABCD} the expressions for
the ${\rm USp}(4)$ covariant derivatives of these tensors
\bea
D_{\mu}  B &=& -\ft25\, P_{\mu\,\jc\jd}{}^{\ja\jb} B^{\jc\jd}{}_{\ja\jb} 
\;,
\nonumber\\[1ex]
D_{\mu}  B^{\ja\jb}{}_{\jc\jd} &=&
-2BP_{\mu\,\jc\jd}{}^{\ja\jb}
-
P_{\mu\,\jg\jh}{}^{\ja\jb}\!B^{\jg\jh}{}_{\jc\jd}
-P_{\mu\,\jc\jd}{}^{\jg\jh}\!B^{\ja\jb}{}_{\jg\jh}+
\ft25
(\delta^{\ja\jb}_{\jc\jd} -\ft14\Omega^{\ja\jb}\Omega_{\jc\jd} ) 
             P_{\mu\,\jg\jh}{}^{\je\jf}\!B^{\jg\jh}{}_{\je\jf}
\;,
\nonumber\\[1ex]
D_{\mu}  C^{\ja\jb} &=& 
   \ft12\, P_{\mu\,\jc\jd}{}^{\ja\jb}A^{\jc\jd} 
    +2\, \Omega^{\je[\ja} P_{\mu\,\jc\jd}{}^{\jb]\jf}
 C^{\jc\jd}{}_{\je\jf} 
\;,
\nonumber\\[1ex]
D_{\mu} C^{\ja\jb}{}_{\jc\jd} &=&
    4\,\Omega^{\jg[\ja}\, P_{\mu\,\jg(\jc}{}^{\jb]\jh} C_{\jd)\jh}
   + \Omega^{\jg[\ja}_{\phantom{(\jc}}\,\delta^{\jb]}_{(\jc}\,
   P_{\mu\,\jd)\jg}{}^{\jk\jh}  C_{\jk\jh}
    +\Omega^{\jg\jk}\,\delta^{[\ja}_{(\jc}\,P_{\mu\,\jd)\jg}{}^{\jb]\jh} C_{\jk\jh}
\nonumber\\[.5ex]
&&{}
+
P_{\mu\,\jg\jh}{}^{\ja\jb} C^{\jg\jh}{}_{\jc\jd}
+
P_{\mu\,\jg\jh}{}^{\jk[\ja}\delta^{\jb]}_{(\jc}\,C^{\jg\jh}{}_{\jd)\jk}
\nonumber\\[.5ex]
&&
{}
+4\,P_{\mu\,\jl(\jc}{}^{\jk\jm}\Omega_{\jd)\jk}\,\Omega^{\jn[\ja}\,C^{\jb]\jl}{}_{\jm\jn}
- \delta^{[\ja}_{(\jc}\,\Omega^{\phantom{\ja}}_{\jd)\jk}\,\Omega^{\jb]\jn}\,
P_{\mu\,\jl\jg}{}^{\jk\jm}C^{\jg\jl}{}_{\jm\jn}
\;.
\eea
Since the $T$-tensor~(\ref{TT}) is obtained
by a finite ${\rm SL}(5)$-transformation
from the embedding tensor~(\ref{linear}),
the ${\rm SL}(5)$-covariant quadratic constraints~(\ref{quadratic})
directly translate into quadratic relations 
among the tensors $B$, $C$.
E.g. the first equation of~(\ref{Q2})
gives rise to
\bea
Z^{(\ja\jb)[\je\jf]}\,
\Big[ 
\Omega_{\jc\je} \Omega_{\jd\jf} \,B
+ \Omega_{\je\jg} \Omega_{\jf\jh} {B^{[\jg\jh]}}_{[\jc\jd]} \Big]
&=& 0
\;,
\label{QZB1}
\eea
while the second equation yields
\bea
Z^{(\ja\jb)[\jc\jd]}\,T_{(\ja\jb)\,\je\jf}{}^{\jg\jh} &=& 0
\;.
\label{QZB2}
\eea
These equations can be further expanded into explicit
quadratic relations among the tensors $B$, $C$.
We give the explicit formulas in terms of ${\rm USp}(4)$ projectors
in appendix~\ref{app:TQC}. They are crucial to verify the
invariance of the Lagrangian~\eqref{L} presented in the next section.

Let us close this section by noting that the $T$-tensor~(\ref{TT})
naturally appears in the deformation of the Cartan-Maurer equations
induced by the gauging. Namely, the definition of the 
currents~$P_\mua$ and $Q_\mua$~\eqref{DefPQ} together with the 
algebra structure~(\ref{SplitCom}) gives rise to the following
integrability relations
\begin{align}
 2 \partial_{[\mua} {Q_{\mub]\ja}}^\jb + 2 {Q_{\ja[\mua}}^\jc {Q_{\mub]\jc}}^\jb
                             &= - 2 {P_{\ja\jc[\mua}}^{\jd\je} {P_{\mub]\jd\je}}^{\jb\jc}
                             - g \, {\cal H}_{\mua\mub}^{(2)\,\jc\jd} \, {T_{(\jc\jd)[\ja\je]}}^{[\jb\je]} 
                             \;,\label{Cartan-Maurer}
\\[1ex]
   D_{[\mua} {P_{\mub]\ja\jb}{}^{\jc\jd}} &= 
   - \; \frac 1 4 g \, {\cal H}_{\mua\mub}^{(2)\,\je\jf} 
                           \left( {T_{(\je\jf)[\ja\jb]}}^{[\jc\jd]} + 
        \Omega^{\jc\jg} \Omega^{\jd\jh} \Omega_{\ja\ji} \Omega_{\jb\jj} \; 
        {T_{(\je\jf)[\jg\jh]}}^{[\ji\jj]} \right) \; .
\nonumber
\end{align}
The terms in order $g$ occur proportional to the 
$T$-tensor. They will play an important role in the check of
supersymmetry of the Lagrangian that we present in the next section.
The fact that these equations appear manifestly
covariant with the full modified field strength ${\cal H}_{\mua\mub}^{(2)\,\jc\jd}$
on the r.h.s.\ is a consequence of 
the quadratic constraint~(\ref{QZB2}).

\section{Lagrangian and supersymmetry}
\setcounter{equation}{0}
\label{SecLagrSUSY}

In this section we present the main results of this paper. After establishing
our spinor conventions, we derive the supersymmetry transformations
of the seven-dimensional theory by requiring closure of
the supersymmetry algebra into the generalized
vector/tensor gauge transformations introduced in section~\ref{SecVecTen}.
We then present the universal Lagrangian of the maximal seven-dimensional theory
which is completely encoded in the embedding tensor $\Theta$.

\subsection{Spinor conventions}
\label{SubSecSpinor}

Seven-dimensional world and tangent-space indices are denoted by
$\mu,\nu,\ldots$ and $m,n,\ldots$, respectively, and take the values
$1,2,\ldots,7$. Our conventions for the $\Gamma$-matrices in seven dimensions are
\begin{align}
   \left\{  \Gamma^{\ma}, \Gamma^{\mb}  \right\} &= 2 \eta^{\ma\mb} &
   (\Gamma^\ma)^\dag &= \Gamma_\ma \; ,&(\Gamma^\ma)^T  &=-C \Gamma^\ma C^{-1} 
\end{align}
with metric of signature $\eta=\text{diag}(-1,1,1,1,1,1,1) $ 
and the charge conjugation matrix $C$ obeying   
\begin{align}   
   C &= C^T = - C^{-1} = - C^\dag \; .
\end{align}
We use symplectic Majorana spinors, i.e.\
spinors carry  a fermionic representation of the $R$-symmetry group ${\rm USp}(4)$
and for instance a spinor  $\psi^\ja$ ($\ja=1, \ldots, 4$) in the fundamental representation 
of ${\rm USp}(4)$ satisfies a reality constraint of the form
\begin{align}
   {\bar \psi_{\ja}}^T &= \Omega_{\ja\jb} C \,\psi^\jb \;, 
   \label{symM}  
\end{align}
where $\bar \psi\equiv\psi^\dag \Gamma^0$.
The following formula is useful as it captures the symmetry property of spinor 
products\footnote{Note that our conventions differ from those of \cite{Sezgin:1982gi}
in that they use ${\phi}_\ja=\Omega_{\ja\jb}{\phi}^\jb$,  while 
in our conventions raising and lowering of indices is
effected by complex conjugation ${\phi}_\ja = ({\phi}^\ja)^*$.}
\begin{align}
   \bar \phi_\ja \Gamma^{(k)} \psi^\jb &= 
   \Omega_{\ja\jc} \Omega^{\jb\jd} \bar \psi_\jd (C^{-1})^T (\Gamma^{(k)})^T C \phi^\jc 
   ~=~ (-1)^{\frac12k(k+1)}\,\Omega_{\ja\jc} \Omega^{\jb\jd} \bar \psi_\jd\Gamma^{(k)} \phi^\jc  \; .
\end{align}
Products of symplectic Majorana spinors yield real tensors
\begin{align}
   \bar \phi_\ja \psi^\ja &&
   \bar \phi_\ja \Gamma^\mua \psi^\ja &&
   \bar \phi_\ja \Gamma^{\mua\mub} \psi^\ja &&
   \bar \phi_\ja \Gamma^{\mua\mub\muc} \psi^\ja &&
   \text{etc.}
\end{align}     
Finally, the epsilon tensor is defined by
\begin{align}
   e\,\Gamma^{\mu\nu\rho\sigma\tau\kappa\lambda} &\equiv \mathbbm{1} \; 
   \epsilon^{\mu\nu\rho\sigma\tau\kappa\lambda} \; .
\end{align}

\subsection{Supersymmetry transformations and algebra}

The field content of the ungauged maximal supergravity multiplet in seven dimensions is
given by the vielbein ${e_\mua}^\ma$, the gravitino $\psi^\ja_\mua$,
vector fields $A_\mua^{\Ja\Jb}$, two-form fields $B_{\Ja \, \mua\mub}$, 
matter fermions $\chi^{\ja\jb\jc}$, and scalar fields parametrizing 
${{\cal V}_\Ja}^{\ja\jb}$.
Their on-shell degrees of freedom are summarized in Table~\ref{TabMultiplet}.
Note the symmetry in the distribution of degrees of freedom
due to the accidental coincidence of the $R$-symmetry group 
${\rm USp}(4)$ and the little group ${\rm SO}(5)$.

\begin{table}[tb]
   \begin{center}
     \begin{tabular}{r|cccccc}
        fields & ${e_\mua}^\ma$ & $\psi^\ja_\mua$ & $A_\mua^{\Ja\Jb}$ & $B_{\mua\mub\,\Ja}$ 
& $\chi^{\ja\jb\jc}$ & ${{\cal V}_\Ja}^{\ja\jb}$
\\ \hline
little group ${\rm SO}(5)$    & {\bf 14} & {\bf 16} & {\bf 5}  & {\bf 10} & {\bf 4} & {\bf 1} \\
$R$-symmetry ${\rm USp}(4)$ & {\bf 1}  & {\bf 4}  
& {\bf 1} & {\bf 1}  & {\bf 16} & {\bf 5}
\\
global ${\rm SL}(5)$ & {\bf 1}  & {\bf 1}  
& ${\bf \overline{10}}$ & {\bf 5}  & {\bf 1} & {\bf 5}
        \\ \hline
        \# degrees of freedom & 14 & 64 & 50 & 50 & 64 & 14 
     \end{tabular}
     \caption{\label{TabMultiplet}{ \small The ungauged $D=7$ maximal supermultiplet.}}
   \end{center}     
\end{table}

Under the $R$-symmetry group ${\rm USp}(4)$ 
the gravitinos $\psi_{\mu}^{\ja}$ 
transform in the fundamental representation ${\bf 4}$
while the matter spinors $\chi^{\ja\jb\jc}$ 
transform in the ${\bf 16}$ representation, i.e.
\begin{align}
   \chi^{\ja\jb\jc} &= \chi^{[\ja\jb]\jc}\;, &
   \Omega_{\ja\jb} \chi^{\ja\jb\jc} &= 0\;, &
   \chi^{[\ja\jb\jc]} &= 0 \; .
\end{align}
All spinors are symplectic Majorana, that is they satisfy
\begin{align}   
   {\bar \chi_{\ja\jb\jc}}^T &= \Omega_{\ja\jd} \Omega_{\jb\je} \Omega_{\jc\jf} C \chi^{\jd\je\jf}\;, &
   {\bar \psi_{\mua\ja}}^T &= \Omega_{\ja\jb} C \psi_\mua^\jb \; ,
\end{align}
in accordance with~(\ref{symM}).

We are now in position to derive the supersymmetry transformations.
Parametrizing them by $\epsilon^\ja=\epsilon^\ja(x)$ the final result takes the form
\begin{align}
   \delta {e_\mua}^\ma &= \frac 1 2 \bar \epsilon_\ja \Gamma^\ma \psi^\ja_\mua 
   \;,
   \nonumber \\[1ex]
   \delta {{\cal V}_\Ja}^{\ja\jb} &= \frac{1}{4} {{\cal V}_\Ja}^{\jc\jd} 
      \Big( \Omega_{\je[\jc} \bar \epsilon_{\jd]} \chi^{\ja\jb\je} 
            + \frac 1 4 \Omega_{\jc\jd} \bar \epsilon_\je \chi^{\ja\jb\je} 
+ \Omega_{\jc\je} \Omega_{\jd\jf} \bar \epsilon_\jg \chi^{\je\jf[\ja} \Omega^{\jb]\jg}
+ \frac 1 4 \Omega_{\jc\je} \Omega_{\jd\jf} \Omega^{\ja\jb} \bar \epsilon_\jg \chi^{\je\jf\jg} \Big) 
\;,
\nonumber \\[1ex]
   \Delta A_\mua^{\Ja\Jb} &= - {{\cal V}_{\ja\jb}}^{[\Ja} {{\cal V}_{\jc\jd}}^{\Jb]} \Omega^{\jb\jd} 
                             \Big( \frac{1}{2} \Omega^{\ja\je} \bar \epsilon_\je \psi_\mua^\jc
+\frac{1}{4} \bar \epsilon_\je \Gamma_\mua \chi^{\je\ja\jc}  \Big)
\;,
\nonumber  \\[1ex]
   \Delta B_{\mua\mub\,\Ja} &= {{\cal V}_{\Ja}}^{\ja\jb}
                             \Big( - \Omega_{\ja\jc} \bar \epsilon_\jb \Gamma_{[\mua} \psi^\jc_{\mub]}
+\frac{1}{8}  \Omega_{\ja\jc} \Omega_{\jb\jd} \bar \epsilon_\je \Gamma_{\mua\mub} \chi^{\jc\jd\je} \Big) 
\;,
\nonumber  \\[1ex]
   \Delta S_{\mua\mub\muc}^\Ja &= {{\cal V}_{\ja\jb}}^{\Ja} 
                           \Big( -\frac{3}{8} \Omega^{\ja\jc} \bar \epsilon_\jc \Gamma_{[\mua\mub} \psi^\jb_{\muc]}
-\frac{1}{32} \bar \epsilon_\je \Gamma_{\mua\mub\muc} \chi^{\ja\jb\je} \Big)  
\;,
\nonumber \\[1ex]
   \delta \psi_\mua^\ja &= D_\mua \epsilon^\ja 
            -\frac{1}{5 {\sqrt{2}}} {\cal H}^{(2)(\ja\jb)}_{\mub\muc} \Omega_{\jb\jc} 
\Big( {\Gamma^{\mub\muc}}_\mua + 8 \Gamma^\mub \delta^\muc_\mua \Big) \epsilon^\jc
            \nonumber \\ & \qquad 
            -\frac{1}{15} {\cal H}^{(3)}_{\mub\muc\mud[\jb\jc]} \Omega^{\ja\jb}     
\Big( {\Gamma^{\mub\muc\mud}}_\mua + \frac 9 2 \Gamma^{\mub\muc} \delta^\mud_\mua \Big) \epsilon^\jc
            - g \Gamma_\mua A_1^{\ja\jb} \Omega_{\jb\jc} \epsilon^\jc 
\;,
\nonumber \\[1ex]
   \delta \chi^{\ja\jb\jc} &= 2 \Omega^{\jc\jd} {P_{\mua\jd\je}}^{\ja\jb} \Gamma^\mua \epsilon^\je
               -\sqrt{2} \Big( {\cal H}_{\mua\mub}^{(2)\jc[\ja} \Gamma^{\mua\mub} \epsilon^{\jb]}
                         - \frac 1 5 \, 
                         (\Omega^{\ja\jb}\delta_{\jg}^{\jc}- \Omega^{\jc[\ja}\delta_{\jg}^{\jb]})\,
                         \Omega_{\jd\je}{\cal H}_{\mua\mub}^{(2)\jg\jd} \Gamma^{\mua\mub} \epsilon^\je
                         \Big)
            \nonumber \\ & \qquad 
           -\frac{1}{6} \Big( \Omega^{\ja\jd} \Omega^{\jb\je} {\cal H}^{(3)}_{\mua\mub\muc[\jd\je]} \Gamma^{\mua\mub\muc} \epsilon^{\jc}
                - \frac 1 5 (\Omega^{\ja\jb} \Omega^{\jc\jf}+4\Omega^{\jc[\ja} \Omega^{\jb]\jf})\,
                 {\cal H}^{(3)}_{\mua\mub\muc[\jf\je]} \Gamma^{\mua\mub\muc} \epsilon^\je
                 \Big)
      \nonumber  \\[1ex]
  &\qquad   + g A_2^{\jd,\ja\jb\jc} \Omega_{\jd\je} \epsilon^\je
   \;,
   \label{SUSYrules}
\end{align}
up to higher order fermion terms.
We have given the result in terms of the covariant variations $\Delta({\epsilon})$ of the vector
and tensor fields introduced in \eqref{covV}, from which the bare 
transformations $\delta({\epsilon})$ are readily deduced.
In the limit $g \rightarrow 0$ the above supersymmetry transformations reduce
to those of the ungauged theory~\cite{Sezgin:1982gi}.
Upon switching on the gauging, the formulas are covariantized
and the fermion transformations are modified by
the fermion shift matrices $A_1$ and $A_2$ defined by
\begin{align}
   A_1^{\ja\jb} &\equiv - \frac{1}{4 {\sqrt{2}}} 
   \Big( \frac 1 4 B \Omega^{\ja\jb} + \frac{1}{5} C^{\ja\jb} \Big) 
\;,
     \nonumber \\
   A_2^{\jd,\ja\jb\jc} &\equiv \frac{1}{2 {\sqrt{2}}} 
   \left[\Omega^{\je\jc} \Omega^{\jf\jd}\,({C^{\ja\jb}}_{\je\jf}- {B^{\ja\jb}}_{\je\jf})
+ \frac{1}{4} \Big( C^{\ja\jb} \Omega^{\jc\jd}
+ \frac 1 5 \Omega^{\ja\jb} C^{\jc\jd} 
+ \frac 4 5 \Omega^{\jc[\ja} C^{\jb]\jd} \Big) \right]   \; ,
   \label{FermionShifts}       
\end{align}
in terms of the components of the $T$-tensor~(\ref{TABCD}).
These will further enter the fermionic mass matrices
and the scalar potential of the full Lagrangian~(\ref{L}) below.
The coefficients in~(\ref{SUSYrules})
are uniquely fixed by requiring the closure of the
supersymmetry algebra
into diffeomorphisms, local Lorentz and ${\rm USp}(4)$-transformations,
and vector/tensor gauge transformations~(\ref{gauge}).
In particular, the fermion shifts~\eqref{FermionShifts}
are uniquely determined such that the 
commutator of two supersymmetry transformations
reproduces the correct order $g$ shift terms 
in the resulting vector/tensor gauge transformations~(\ref{gauge}).
Specifically, one finds for the commutator of two supersymmetry transformations
\begin{align}
   [ \delta(\epsilon_1), \delta(\epsilon_2) ] &=  \xi^\mua D_\mua  
                                           + \delta_{\text{Lorentz}}  \left(\epsilon^{\ma\mb} \right)
+ \delta_{{\rm USp}(4)} \left( {\kappa_\ja}^\jb \right)
+ \delta_{\text{gauge}}   \Big( \Lambda^{\Ja\Jb}, \Xi_{\Ja\mua}, \Phi^\Ja_{\mua\mub} \Big) \; .
   \label{ComSUSYalgebra}      
\end{align}
Here, we denote by $\xi^\mua D_\mua$ a covariant general coordinate
transformation with parameter $\xi^\mua$, i.e.
\begin{align}
   \xi^\mua D_\mua &= {\cal L}_\xi + \delta_{\text{Lorentz}}  \left( \hat \epsilon{}^{\;\ma\mb} \right)
+ \delta_{{\rm USp}(4)} \left( {{\hat \kappa}_\ja}{}^\jb \right)
+ \delta_{\text{gauge}}   \left( \hat \Lambda^{\Ja\Jb}, \hat \Xi_{\Ja\mua}, \hat \Phi^\Ja_{\mua\mub} \right) 
\; ,
\end{align}
with the induced parameters
\begin{align}
   \hat \epsilon{}^{\;\ma\mb} &= - \xi^\mua {\omega_\mua}^{\ma\mb} \;,\nonumber \\
   {{\hat \kappa}_\ja}{}^\jb &= - \xi^\mua {Q_{\mua\ja}}^\jb\;, \nonumber \\
   \hat \Lambda^{\Ja\Jb} &= - \xi^\mua A_\mua^{\Ja\Jb} \;,\nonumber \\
   \hat \Xi_{\Ja\mua} &= - \xi^\mub B_{\Ja\,\mub\mua}
   -  \epsilon_{\Ja\Jb\Jc\Jd\Je}\xi^\mub A_\mub^{\Jb\Jc}A_\mua^{\Jd\Je}\;,\nonumber \\
   \hat \Phi^\Ja_{\mua\mub} &= - \xi^\muc S^\Ja_{\muc\mua\mub}
   -\xi^{\muc} A_{\muc}^{\Ja\Jb} B_{\mua\mub\,\Jb}
   -\frac23\epsilon_{\Jb\Jc\Jd\Je\Jf}\,\xi^{\muc} A_{\muc}^{\Jb\Jc}
   A_{[\mua}^{\Ja\Jd}A_{\mub]}^{\vphantom{\Jd}\Je\Jf}
     \; .
\end{align}
In addition to these transformations the right hand side of \eqref{ComSUSYalgebra}
consists of general coordinate, Lorentz, ${\rm USp}(4)$, and
vector/tensor gauge transformations with parameters given by
\begin{align}
   \xi^\mua &= \frac 1 2 \bar \epsilon_{2\ja} \Gamma^\mua \epsilon_1^\ja \;,
       \nonumber \\
   \epsilon^{\ma\mb} &= -\frac{1}{5 {\sqrt{2}}} {\cal H}^{(2)(\ja\jb)}_{\mc\md} \Omega_{\jb\jc} 
                    \bar \epsilon_{2\ja} \left( \Gamma^{\ma\mb\mc\md} + 
                    8 \eta^{\ma\mc} \eta^{\mb\md} \right) \epsilon_1^\jc 
                    + \frac{g}{20 {\sqrt{2}}} A^{\ja\jb} \Omega_{\jb\jc} \bar \epsilon_{2\ja} \Gamma^{\ma\mb} \epsilon_1^\jc 
\nonumber \\ & \qquad 
                  + \frac{1}{15} {\cal H}^{(3)}_{\mc\md\me[\ja\jb]} \Omega^{\jb\jc}
\bar \epsilon_{2\jc} \left( \Gamma^{\ma\mb\mc\md\me} + 
9 \eta^{\ma\mc} \eta^{\mb\md} \Gamma^\me \right) \epsilon_1^\ja
- \frac{g}{16 {\sqrt{2}}} D \bar \epsilon_{2\ja} \Gamma^{\ma\mb} \epsilon_1^\ja   
       \;,\nonumber \\
   {\kappa_\ja}^\jb &= \frac 1 4 \Lambda^{\jd\je} {T_{(\jd\je)[\ja\jc]}}^{[\jb\jc]} \;,
       \nonumber \\
   \Lambda^{\Ja\Jb} &= \sqrt{2} {{\cal V}_{\ja\jb}}^\Ja {{\cal V}_{\jc\jd}}^\Jb \Omega^{\jb\jd} \Lambda^{\ja\jc} \; ,
       \qquad \qquad 
   \text{with} \qquad  \Lambda^{\ja\jb}  = \frac{1}{2\sqrt{2}} \Omega^{\jc(\ja} \bar \epsilon_{2\jc} \epsilon_1^{\jb)} 
     \;,   \nonumber \\
   \Xi_{\Ja\mua} 
      &= \frac {1} 2 {{\cal V}_\Ja}^{\ja\jb} \bar \epsilon_{2\ja} \Gamma_\mua \epsilon_1^\jc \Omega_{\jc\jb} 
   \;,     \nonumber \\
   \Phi^\Ja_{\mua\mub} &= - \frac{1} 8 {{\cal V}_{\ja\jb}}^\Ja \Omega^{\ja\jc} 
                               \bar \epsilon_{2\jc} \Gamma_{\mua\mub} \epsilon_1^\jb
                               \;.
\end{align}
To this order in the fermion fields the fermionic field equations are not
yet required for verifying the closure~(\ref{ComSUSYalgebra}) of the algebra.
Closure on the three-form tensor fields 
$S^\Ja_{\mua\mub\muc}$ however makes use of the (projected) duality equation
\bea
   e^{-1} \, \epsilon^{\mua\mub\muc\mud\mue\muf\mug} Y_{\Ja\Jb} 
   {\cal H}^{(4)\Jb}_{\mud\mue\muf\mug}
   &=& 6 \, Y_{\Ja\Jb} \, \Omega^{\ja\jc} \Omega^{\jb\jd} \, 
   {{\cal V}_{\ja\jb}}^\Jb {\cal H}^{(3)}_{\jc\jd \, \mua\mub\muc}
     + \text{fermionic terms}\;,
   \label{Duality1}
\eea
between two- and three form tensor fields.
{}From~(\ref{VarHH}), (\ref{LkinH3}) and (\ref{varCS})
one may already anticipate that this equation 
will arise as a first order equation of motion from
the full Lagrangian upon varying w.r.t.\ the $S^{\Ja}_{\mua\mub\muc}$.
We will confirm this in the next section.
Note that also this duality equation appears only under 
projection with $Y_{\Ja\Jb}$.

\subsection{The universal Lagrangian}

We can now present the universal Lagrangian of gauged maximal supergravity 
in seven dimensions up to higher order fermion terms:
\begin{align}
   e^{-1} {\cal L} =& 
               - \frac 1 2 R 
-  \Omega_{\ja\jc} \Omega_{\jb\jd} {\cal H}^{(2)\ja\jb}_{\mua\mub} {\cal H}^{(2)\jc\jd\mua\mub}
- \frac 1 {6} \Omega^{\ja\jc} \Omega^{\jb\jd} {\cal H}^{(3)}_{\mua\mub\muc\,\ja\jb} 
{\cal H}^{(3)}{}_{\jc\jd}^{\mua\mub\muc}
- \frac 1 2 {P_{\mua\ja\jb}}^{\jc\jd} {P^\mua}_{\jc\jd}{}^{\ja\jb}
\nonumber \\ & 
- \frac 1 2 \bar \psi_{\mua\ja} \Gamma^{\mua\mub\muc} D_\mub \psi_{\muc}^\ja
- \frac 1 {8} \bar \chi_{\ja\jb\jc} \slashchar D \chi^{\ja\jb\jc}
- \frac 1 2 {P_{\mua\ja\jb}}^{\jc\jd} \Omega_{\jc\je} \bar \psi_{\mub\jd} \Gamma^{\mua} \Gamma^{\mub} \chi^{\ja\jb\je}
\nonumber \\ & 
+ \frac{\sqrt{2}} {4} {\cal H}^{(2)\ja\jb}_{\mua\mub} \,
\Big( - \bar \psi^\muc_{\ja} \Gamma_{[\muc} \Gamma^{\mua\mub} \Gamma_{\mud]} \psi^{\mud\jc} \Omega_{\jc\jb} 
                         + \bar \psi_{\muc\jc} \Gamma^{\mua\mub} \Gamma^{\muc} \chi^{\jc\jd\je} \Omega_{\ja\jd} \Omega_{\jb\je}
+\frac 1 2 \bar \chi_{\ja\jc\jd} \Gamma^{\mua\mub} \chi^{\je\jd\jc} \Omega_{\je\jb}  \Big)
\nonumber \\ & 
               + \frac 1 {12} {\cal H}^{(3)}_{\ja\jb\mua\mub\muc} \,
\Big( -\Omega^{\ja\jc} \bar \psi^\mud_{\jc} \Gamma_{[\mud} \Gamma^{\mua\mub\muc} \Gamma_{\mue]} \psi^{\mue\jb}
                         +\frac 1 2 \bar \psi_{\mud\jc} \Gamma^{\mua\mub\muc} \Gamma^{\mud} \chi^{\ja\jb\jc}
+\frac 1 4 \Omega^{\ja\je} \bar \chi_{\jc\jd\je} \Gamma^{\mua\mub\muc} \chi^{\jc\jd\jb} \Big)
\nonumber \\ &  
               - \frac 5 2 g A_1^{\ja\jb} \Omega_{\jb\jc} \bar \psi_{\mua\ja} \Gamma^{\mua\mub} \psi^{\jc}_\mub
+ \frac 1 4 g A_2^{\jd,\ja\jb\jc} \Omega_{\jd\je} \bar \chi_{\ja\jb\jc} \Gamma^{\mua} \psi_\mua^{\je}
\nonumber \\ &  
+ \frac{g} {4 \sqrt{2}} \, \Big( \frac 3 {32} \delta^\jb_\jd \delta^\jc_\je B
+\frac 1 8 \delta^\jb_\jd \Omega_{\je\jf} C^{\jf\jc}
+ {B^{\jb\jc}}_{\jd\je} - {C^{\jb\jc}}_{\jd\je} \Big) 
\bar \chi_{\ja\jb\jc} \, \chi^{\ja\jd\je}
\nonumber \\ & 
+ \frac {g^2} {128} \left(15 B^2+2 C^{\ja\jb}C_{\ja\jb} - 
2 B^{\ja\jb}{}_{\jc\jd}B^{\jc\jd}{}_{\ja\jb} - 
2 C^{[\ja\jb]}{}_{(\jc\jd)}C_{[\ja\jb]}{}^{(\jc\jd)}
 \right) 
\nonumber\\[1ex]
&+ e^{-1}{\cal L}_{\rm VT}
\;,
\label{L}
\end{align}
with the tensors $A_{1}$, $A_{2}$ from~(\ref{FermionShifts}) and the
topological vector-tensor Lagrangian from (\ref{VT}): 
\begin{align*}
{\cal L}_{\rm VT} &= - \frac 1 {9}
\epsilon^{\mua\mub\muc\mud\mue\muf\mug}\;\times
\nonumber\\
&
\times     
\Big[ g Y_{\Ja\Jb} S^\Ja_{\mua\mub\muc} \Big( D^{\vphantom{\Ja}}_\mud S_{\mue\muf\mug}^\Jb
+ \frac 3 2 g Z^{\Jb\Jc,\Jd} B_{\mud\mue\,\Jc} B_{\muf\mug\,\Jd}
+ 3 {\cal F}^{\Jb\Jc}_{\mud\mue} B^{\vphantom{\Ja}}_{\muf\mug\,\Jc}
\nonumber \\ & \qquad \qquad
                 + 4 \epsilon_{\Jc\Jd\Je\Jf\Jg} A^{\Jb\Jc}_{\mud} A^{\Jd\Je}_{\mue} \partial^{\vphantom{\Ja}}_{\muf} A_\mug^{\Jf\Jg}
        + g \epsilon_{\Jc\Jd\Je\Jj\Jk} {X_{\Jf\Jg,\Jh\Ji}}^{\Jj\Jk} A_\mud^{\Jb\Jc} A_\mue^{\Jd\Je} A_\muf^{\Jf\Jg} A_\mug^{\Jh\Ji} \Big) 
  \nonumber \\  &  \quad
      + 3 g Z^{\Ja\Jb,\Jc} (D_\mua B_{\mub\muc\,\Ja}) B_{\mud\mue\,\Jb} B_{\muf\mug\,\Jc} 
    - \frac 9 2 {\cal F}_{\mua\mub}^{\Ja\Jb} B_{\muc\mud\,\Ja} D_{\mue} B_{\muf\mug\,\Jb}
  \nonumber \\ & \quad
    + 18 \epsilon_{\Ja\Jb\Jc\Jd\Je} {\cal F}_{\mua\mub}^{\Ja\Ji} A^{\Jb\Jc}_{\muc} 
            \Big( \partial^{\vphantom{\Ja}}_\mud A^{\Jd\Je}_{\mue} 
                                       + \frac 2 3 g {X_{\Jf\Jg,\Jh}}^\Jd A^{\Je\Jh}_\mud A^{\Jf\Jg}_{\mue} \Big) B_{\muf \mug\,\Ji}
  \nonumber \\ & \quad
    + 9 g \epsilon_{\Ja\Jb\Jc\Jd\Je} Z^{\Ja\Ji,\Jj} A^{\Jb\Jc}_{\mua} \Big( \partial^{\vphantom{\Ja}}_\mub A^{\Jd\Je}_{\muc} 
          + \frac 2 3 g {X_{\Jf\Jg,\Jh}}^\Jd A^{\Je\Jh}_\mub A^{\Jf\Jg}_{\muc} \Big) B_{\mud\mue\,\Ji} B_{\muf\mug\,\Jj} \nonumber \\
  & + \frac {36} 5 \epsilon_{\Ja\Jc\Jd\Jg\Jh} \; \epsilon_{\Jb\Je\Jf\Ji\Jj} 
       A_\mua^{\Ja\Jb} A_\mub^{\Jc\Jd} A_\muc^{\Je\Jf} 
             (\partial_\mud^{\vphantom{\Ja}} A_\mue^{\Jg\Jh}) (\partial_\muf^{\vphantom{\Ja}} A_\mug^{\Ji\Jj})
     \nonumber \\ &   
    + 8 g \epsilon_{\Ja\Jc\Jd\Je\Jf} \; \epsilon_{\Jb\Jg\Jh\Jm\Jn} {X_{\Ji\Jj,\Jk\Jl}}^{\Jm\Jn}
          A_\mua^{\Ja\Jb} A_\mub^{\Jc\Jd} A_\muc^{\Jg\Jh} A_\mud^{\Ji\Jj} A_\mue^{\Jk\Jl} 
\partial_\muf^{\vphantom{\Ja}} A_\mug^{\Je\Jf} \nonumber \\
  & - \frac 4 7 g^2 \epsilon_{\Ja\Jc\Jd\Jo\Jp} \; \epsilon_{\Jb\Ji\Jj\Jq\Jr} {X_{\Je\Jf,\Jg\Jh}}^{\Jo\Jp} {X_{\Jk\Jl,\Jm\Jn}}^{\Jq\Jr}
        A_\mua^{\Ja\Jb} A_\mub^{\Jc\Jd} A_\muc^{\Je\Jf} A_\mud^{\Jg\Jh} A_\mue^{\Ji\Jj} A_\muf^{\Jk\Jl} A_\mug^{\Jm\Jn}
\Big]
\;.
\end{align*}
This Lagrangian is the unique one invariant under the full set of 
nonabelian vector/tensor gauge 
transformations~\eqref{gauge} and under local 
supersymmetry 
transformations~\eqref{SUSYrules}.
Furthermore it possesses the local ${\rm USp}(4)$ invariance 
introduced in~\eqref{TrafoCalV}, and is formally
invariant under global ${\rm SL}(5)$ transformations
if the embedding tensor~$\Theta$ is treated as a spurionic
object that simultaneously transforms. 
With fixed~$\Theta$, the global ${\rm SL}(5)$
is broken down to the gauge group.

In the limit $g \rightarrow 0$ the three-form fields $S^{\Ja}_{\mua\mub\muc}$
decouple from the Lagrangian, and~(\ref{L}) consistently 
reduces to the  ungauged theory of~\cite{Sezgin:1982gi} with global ${\rm SL}(5)$ 
symmetry.
Upon effecting the deformation by switching on $g$,
derivatives are covariantized $\partial_\mua \rightarrow D_\mua$ 
and the former abelian field strengths are replaced by the full covariant
combinations ${\cal H}^{(2)}$ and ${\cal H}^{(3)}$ from~(\ref{H2}), (\ref{H3}).
As discussed in section~4, the extended gauge invariance~\eqref{gauge} 
moreover requires a unique extension of the former abelian topological
term which in particular includes a first order kinetic term for the 
three-form fields $S^{\Ja}_{\mua\mub\muc}$.
As a consequence, the duality equation \eqref{Duality1} 
between the two-form and the three-form
tensor fields arises directly as a field equation of this Lagrangian. 
This ensures that the total number of degrees of freedom
is not altered by switching on the deformation
and does not depend on the explicit form of 
the embedding tensor.

In order to maintain supersymmetry under the extended 
transformations~\eqref{SUSYrules}, 
and in presence of the deformed Bianchi 
and Cartan-Maurer equations~\eqref{Bianchi}, 
\eqref{Cartan-Maurer}, the Lagrangian
finally needs to be augmented by the bilinear fermionic mass terms 
in order $g$ and a scalar potential in order $g^{2}$. 
These are expressed in terms of the scalar field dependent 
${\rm USp}(4)$-components $B$, $C$ of the $T$-tensor.
Cancellation of the terms in order $g^{2}$ in particular
requires the quadratic identities~\eqref{QZB1}, \eqref{QZB2},
expanded in components in~\eqref{Con5}, \eqref{Con35}. 
In particular, these identities give rise to
\bea
\frac18 A_2^{\ja,\jc\jd\je}A^{\vphantom{\jc}}_{2\,\jb,\jc\jd\je}-15 
A_{1}^{\ja\jc}A^{\vphantom{\jc}}_{1\,\jb\jc} &=&
\frac14\delta_{\jb}^{\ja}\,
\Big(
\frac18 A_2^{\jf,\jc\jd\je}A^{\vphantom{\jc}}_{2\,\jf,\jc\jd\je}-15 
A_{1}^{\jc\jd}A^{\vphantom{\jc}}_{1\,\jc\jd}
\Big)\;,
   \label{A1A1A2A2}
\eea
featuring the scalar potential on the r.h.s.\ and needed
for cancellation of the supersymmetry contributions from the scalar potential.
Indeed, the scalar potential which contributes to the 
Lagrangian~\eqref{L} in order $g^2$ may be written in the equivalent forms
\begin{align}
   V &= - \frac {1} {128} \left(15 B^2+2 C^{\ja\jb}C_{\ja\jb} - 
2 B^{\ja\jb}{}_{\jc\jd}B^{\jc\jd}{}_{\ja\jb} - 
2 C^{[\ja\jb]}{}_{(\jc\jd)}C_{[\ja\jb]}{}^{(\jc\jd)}
 \right) 
 \nonumber\\[1ex] & 
 = \frac 1 8 |A_2|^2 - 15 |A_1|^2 
 \;.
   \label{SPotential}
\end{align}
Under variation of the scalar fields given by 
$\delta_{\Sigma} {\cal V}_{M}{}^{ab}=\Sigma^{ab}{}_{cd}\,{\cal V}_{M}{}^{cd}$ 
the potential varies according to
\bea
  \delta_{\Sigma} V &= &
          -\frac {1}{16} {B^{[\ja\jb]}}_{[\jc\jd]} 
          {B^{[\jc\jd]}}_{[\je\jf]} {\Sigma^{[\je\jf]}}_{[\ja\jb]}
+\frac {1}{32} B {B^{[\ja\jb]}}_{[\jc\jd]} {\Sigma^{[\jc\jd]}}_{[\ja\jb]}
          -\frac {1}{64} C^{[\ja\jb]} C_{[\jc\jd]} {\Sigma^{[\jc\jd]}}_{[\ja\jb]}
\nonumber \\[1ex]  && 
{}
+\frac {1}{32} {C^{[\ja\jb]}}_{(\je\jf)} {C_{[\jc\jd]}}^{(\je\jf)} {\Sigma^{[\jc\jd]}}_{[\ja\jb]}
-\frac{1}8
{C^{[\jc\je]}}_{(\ja\jf)} {C^{[\jd\jf]}}_{(\jb\je)} 
{\Sigma^{[\ja\jb]}}_{[\jc\jd]} \;,
   \label{VaryV}
\eea
which in particular yields the contribution of the potential
under supersymmetry transformations.
Moreover, equation~\eqref{VaryV} 
is important when analyzing the ground states of the theory since
$\delta_{\Sigma} V = 0$ is a necessary condition for a stationary point
of the potential.
The residual supersymmetry of the corresponding solution
(assuming maximally symmetric spacetimes) is parametrized by
spinors~$\epsilon^\ja$ satisfying the condition
\begin{equation}
A_{2\,\ja,\jb\jc\jd}\, \epsilon^\ja = 0 \;.
\label{susy_groundstate}
\end{equation}
The gravitino variation imposes an extra condition
\bea
2A_{1\,\ja\jb}\,\epsilon^{\jb} &=&
\pm \sqrt{-V/15}\,\Omega_{\ja\jb}\,\epsilon^{\jb}
\;,
   \label{susy_groundstate2}
\eea
but the two conditions~(\ref{susy_groundstate})
and~(\ref{susy_groundstate2})
are in fact equivalent by virtue of~(\ref{A1A1A2A2}).\footnote{
More precisely, a solution of~\eqref{susy_groundstate}, (\ref{susy_groundstate2}) 
tensored with a Killing spinor of AdS$_{7}$
(or seven-dimensional Minkowski space, respectively,
depending on the value of $V$)
solves the Killing spinor equations $\delta\psi^{a}_{\mu}=0$,
$\delta\chi^{abc}=0$ obtained from~\eqref{SUSYrules}.
}

The full check of invariance of the Lagrangian~\eqref{L}
under the supersymmetry transformations~\eqref{SUSYrules}
is rather lengthy and makes heavy use
of the quadratic constraints~\eqref{quadratic}
on the embedding tensor and their
consequences collected in appendix~\ref{app:TQC}
as well as of the properties of the ${\rm SL}(5)/{\rm USp}(4)$ coset
space discussed in the previous section.
We have given the Lagrangian and transformation rules only
up to higher order fermion terms; however one does not 
expect any order $g$ corrections to these higher order
fermion terms, i.e.\ they remain unchanged w.r.t.\ those of 
the ungauged theory.

Let us finally note that the bosonic part of the Lagrangian~(\ref{L})
can be cast into a somewhat simpler form in which the scalar fields parametrize
the ${\rm USp}(4)$-invariant symmetric unimodular matrix ${\cal M}_{\Ja\Jb}$
\begin{align}
   {\cal M}_{\Ja\Jb} &\equiv 
   {{\cal V}_\Ja}^{\ja\jb} {{\cal V}_\Jb}^{\jc\jd} \,\Omega_{\ja\jc} \Omega_{\jb\jd}
   \; ,
   \label{DefCalM}
\end{align}
with the inverse ${\cal M}^{\Ja\Jb}= ({\cal M}_{\Ja\Jb})^{-1}=
{{\cal V}_{\ja\jb}}^\Ja {{\cal V}_{\jc\jd}}^\Jb \,\Omega^{\ja\jc} \Omega^{\jb\jd}$. 
The bosonic part of the Lagrangian~(\ref{L}) can then
be expressed exclusively in terms of ${\rm USp}(4)$-invariant quantities
and takes the form
\bea
e^{-1} {\cal L}_{\text{bosonic}} &=&
- \frac 1 2 R 
- {\cal M}_{\Ja\Jc} {\cal M}_{\Jb\Jd} {\cal H}^{(2)\Ja\Jb}_{\mua\mub} {\cal H}^{(2)\mua\mub\,\Jc\Jd}
- \frac 1 6 {\cal M}^{\Ja\Jb} {\cal H}^{(3)}_{\mua\mub\muc\,\Ja} {\cal H}^{(3)}{}^{\mua\mub\muc}_{\Jb}
\nonumber \\ &&
+ \frac 1 8 (\partial_\mua {\cal M}_{\Ja\Jb}) (\partial^\mua {\cal M}^{\Ja\Jb})
 + e^{-1}{\cal L}_{\text{VT}}- g^{2}\,V\;,
 \label{LM}
 \eea
 with the scalar potential 
\bea
V&=&
\frac {1} {64} \Big( 
                         3 {X_{\Ja\Jb,\Je}}^\Jf {X_{\Jc\Jd,\Jf}}^\Je {\cal M}^{\Ja\Jc} {\cal M}^{\Jb\Jd}
-  {X_{\Ja\Jc,\Jd}}^\Jb {X_{\Jb\Je,\Jf}}^\Ja {\cal M}^{\Jc\Je} {\cal M}^{\Jd\Jf}
\Big)
\nonumber\\ && 
+\frac {1} {96} \Big(
 {X_{\Ja\Jb,\Je}}^\Jf {X_{\Jc\Jd,\Jg}}^\Jh 
{\cal M}^{\Ja\Jc} {\cal M}^{\Jb\Jd} {\cal M}^{\Je\Jg} {\cal M}_{\Jf\Jh}
 + {X_{\Ja\Jc,\Jd}}^\Jb {X_{\Jb\Je,\Jf}}^\Ja {\cal M}^{\Jc\Jd} {\cal M}^{\Je\Jf} \Big)
 \nonumber\\[2ex]
 &=&
 \frac1{64}
\Big(2{\cal M}^{\Ja\Jb}Y_{\Jb\Jc}{\cal M}^{\Jc\Jd}Y_{\Jd\Ja}-
({\cal M}^{\Ja\Jb}Y_{\Ja\Jb})^{2}\Big)
\nonumber\\
&&{}
+Z^{\Ja\Jb,\Jc}Z^{\Jd\Je,\Jf}\,\Big(
{\cal M}_{\Ja\Jd}{\cal M}_{\Jb\Je}{\cal M}_{\Jc\Jf}
-{\cal M}_{\Ja\Jd}{\cal M}_{\Jb\Jc}{\cal M}_{\Je\Jf} \Big)
\;.
 \label{LMP}
\eea
This is in analogy to the fact that 
the gravitational degrees of freedom can be described alternatively
in terms of the vielbein or in terms of the metric.
In particular, the scalar potential here is directly expressed 
in terms of the embedding tensor~(\ref{X-theta})
properly contracted with the scalar matrix ${\cal M}$
without having to first pass to the ${\rm USp}(4)$ tensors $B$, $C$.
In concrete examples this may simplify the computation
and the analysis of the scalar potential.
Of course, in order to describe the coupling to fermions
it is necessary to reintroduce ${\cal V}$, the tensors $B$, $C$,  
and to exhibit the local ${\rm USp}(4)$ symmetry.

\section{Examples}
\setcounter{equation}{0}
\label{SecExpamples}

In this section, we will illustrate the general formalism 
with several examples. 
In particular, these include the maximally supersymmetric 
theories resulting from M-theory compactification 
on $S^{4}$~\cite{Pernici:1984xx,Pilch:1984xy,Nastase:1999cb,Nastase:1999kf},
as well as the (warped) type IIA/IIB compactifications on $S^{3}$
which so far have only partially been constructed in the literature.

In order to connect to previous results in the literature, we first discuss 
the possible gauge fixing of tensor gauge transformations
depending on the specific form of the embedding tensor.
In sections~\ref{SecExNoZ} and \ref{SecExNoY} we consider particular
classes of examples in which the embedding tensor is restricted 
to components in either the ${\bf 15}$ or the ${\bf \overline{40}}$
representation. Finally, we sketch in section~\ref{sec:YZ}
a more systematic approach towards classifying the solutions of the 
quadratic constraint~\eqref{quadratic} with both $Y_{\Ja\Jb}$ and $Z^{\Ja\Jb,\Jc}$
nonvanishing. Our findings are collected in Table~\ref{TabSumEx}.

\subsection{Gauge fixing}
\label{sec:gfixing}

We have already noted in section~\ref{SecVecTen}
that the extended local gauge transformations~\eqref{gauge}
allow to eliminate a number of vector and tensor fields
depending on the specific form of the components 
$Y_{\Ja\Jb}$ and $Z^{\Ja\Jb,\Jc}$ of the embedding tensor.
More precisely, $s\equiv{\rm rank}\,Z$ vector fields can be set to 
zero by means of tensor gauge transformations 
$\delta_\Xi$ of~\eqref{gauge},
rendering $s$ of the two-forms massive.
Here, $Z^{\Ja\Jb,\Jc}$ is understood as a rectangular $10\times5$ matrix.
Furthermore, $t\equiv{\rm rank}\,Y$ of the two-forms can be 
set to zero by means of tensor gauge transformations 
$\delta_\Phi$. The $t$ three-forms
that appear in the Lagrangian~\eqref{L} then turn into 
selfdual massive forms.
The quadratic constraint~\eqref{Q2} ensures that $s+t\leq5$.
Before gauge fixing, the degrees of freedom in the Lagrangian~\eqref{L}
are carried by the vector and two-form fields just as in the ungauged theory
(Table~\ref{TabMultiplet}) while the three-forms
appear topologically coupled. After gauge fixing the distribution
of these 100 degrees of freedom is summarized in Table~\ref{dof}.
In a particular ground state, in addition some of the vectors
may become massive by a conventional Brout-Englert-Higgs mechanism.

\begin{table}[bt]
\begin{center}
\begin{tabular}{r|c|c}
fields & $\#$ & $\#$ dof \\
\hline
massless vectors & $10-s$ & 5 \\
massless 2-forms& $5-s-t$ & 10\\
massive 2-forms& $s$ & 15 \\
massive sd.\ 3-forms & $t$ & 10
\end{tabular}
\caption{{\small Distribution of degrees of freedom after gauge fixing.}\label{dof}}
\end{center}
\end{table}

Let us make this a little more explicit. To this end, we employ 
for the two-forms a special basis $B_{\Ja}=(B_{x},B_{\alpha})$,
$x=1,\dots, t$; $\alpha=t\!+\!1, \dots, 5$,
such that the symmetric matrix $Y_{\Ja\Jb}$ takes block diagonal form,
$Y_{xy}$ is invertible (with inverse $Y^{xy}$), 
and all entries $Y_{x\alpha}$, $Y_{\alpha\beta}$ vanish.
For the tensor $Z$ the quadratic constraint~\eqref{quadratic} then implies
that only its components
\bea
Z^{\alpha\beta,\gamma}\;,\qquad 
Z^{x\alpha,\beta}=Z^{x(\alpha,\beta)}
\;,
\label{nZ}
\eea
are nonvanishing and need to satisfy
\bea
Y_{xy}\,Z^{y\alpha,\beta} 
+2\epsilon_{x\Ja\Jb\Jc\Jd}\,Z^{\Ja\Jb,\alpha}Z^{\Jc\Jd,\beta}
&=& 0
\;.
\label{quadraticZ}
\eea
Gauge fixing eliminates the two-forms $B_{x}$ which explicitly breaks the 
${\rm SL}(5)$ covariance.
Supersymmetry transformations thus need to be amended by a compensating term
$\delta^{\rm new}(\epsilon)=\delta^{\rm old}(\epsilon)+
\delta(\Phi_{\mua\mub}^{x})$.
It is convenient to define the modified three-forms
\bea
{\cal S}^{x}_{\mu\nu\rho} &\equiv& 
g^{-1} Y^{xy}\,{\cal H}^{(3)}_{\mu\nu\rho\,y}
~=~  S^{x}_{\mua\mub\muc} + 6 g^{-1} Y^{xy}
\epsilon_{y\Ja\Jb\Jc\Jd} 
A^{\Ja\Jb}_{[\mu}\partial^{\vphantom{\Jc}}_{\nu}A^{\Jc\Jd}_{\rho]}+\dots \;,
\label{newS}
\eea
which are by construction invariant under tensor gauge transformations
and will appear in the Lagrangian as massive fields.
Their transformation under local gauge and supersymmetry 
is easily deduced from~\eqref{trafH3}
\bea
\delta({\Lambda})\,{\cal S}^{x}_{\mu\nu\rho}&=&
-gY_{yz}\Lambda^{xy} {\cal S}^{z}_{\mua\mub\muc} 
-\Lambda^{x\alpha} {\cal H}^{(3)}_{\mua\mub\muc\,\alpha} 
-2 Y^{xy}Z^{\Jb\Jc,\alpha}\,\epsilon_{y\Jb\Jc\Jd\Je}\, 
\Lambda^{\Jd\Je} {\cal H}_{\mua\mub\muc\,\alpha}^{(3)} 
\;,
\nonumber\\[1.5ex]
\delta({\epsilon})\,{\cal S}^{x}_{\mu\nu\rho}&=&
-{{\cal V}_{\ja\jb}}^{x} 
                          ( \ft{3}{8} \Omega^{\ja\jc} \bar \epsilon_\jc \Gamma_{[\mua\mub} \psi^\jb_{\muc]}
+\ft{1}{32} \bar \epsilon_\je \Gamma_{\mua\mub\muc} \chi^{\ja\jb\je})  
\nonumber\\[.5ex]
&&
- 3 g^{-1} Y^{xy}\,\epsilon_{y\Jb\Jc\Jd\Je} {\cal H}_{[\mua\mub}^{(2)\Jb\Jc} 
{{\cal V}_{\ja\jb}}^{[\Jd} {{\cal V}_{\jc\jd}}^{\Je]} \Omega^{\jb\jd} 
                             ( \Omega^{\ja\je} \bar \epsilon_\je \psi_{\muc]}^\jc
+\ft{1}{2} \bar \epsilon_\je \Gamma_{\muc]} \chi^{\je\ja\jc}  )
\nonumber\\[.5ex]
&&
-3g^{-1} Y^{xy}\,  D_{[\mua} \Big(
                             ( \Omega_{\ja\jc} \bar \epsilon_\jb \Gamma^{{\vphantom{]}}}_{\mub} \psi^\jc_{\muc]}
-\ft{1}{8}  \Omega_{\ja\jc} \Omega_{\jb\jd} \bar \epsilon_\je \Gamma_{\mub\muc]} 
\chi^{\jc\jd\je}) {{\cal V}_{y}}^{\ja\jb}\Big) 
\;.
\label{gf1}
\eea
In the Lagrangian these fields appear with a mass term descending from the kinetic
term of the modified field strength tensor
${\cal H}_{\mua\mub\muc\,\ja\jb}=
{\cal V}_{\ja\jb}{}^{\alpha}{\cal H}_{\mua\mub\muc\,\alpha}^{(3)} +
gY_{xy}{\cal V}_{\ja\jb}{}^{x}{\cal S}_{\mua\mub\muc}^{y}$
and a first order kinetic term from the Chern-Simons term
\bea
{\cal L}_{\rm VT} &= - \frac 1 {9}g
\epsilon^{\mua\mub\muc\mud\mue\muf\mug}
 Y_{xy}\, {\cal S}^x_{\mua\mub\muc} D^{\vphantom{\Ja}}_\mud {\cal S}_{\mue\muf\mug}^y
+\dots \;.
\label{gf2}
\eea
The remaining terms in the expansion~\eqref{newS} in particular lead to terms 
$A\,\partial\!A\,\partial\!A\,\partial\!A$
of order $g^{-1}$ in the topological term which obstruct a smooth limit back
to the ungauged theory. Indeed these terms have been observed in the 
original construction of the ${\rm SO}(p,q)$ gaugings~\cite{Pernici:1984xx}.
Generically the gauge fixing procedure described above leads to 
many more interaction terms between vector and tensor fields than those that 
are known from the particular case of the ${\rm SO}(p,q)$ theories.

\subsection{Gaugings in the ${\bf 15}$ representation:~ \\
${\rm SO}(p,5\!-\!p)$ and
${\rm CSO}(p,q,5\!-\!p\!-\!q)$}
\label{SecExNoZ}

As a first class of examples let us analyze those gaugings
for which the embedding tensor $\Theta$ lives entirely in the ${\bf 15}$
representation of ${\rm SL}(5)$, i.e.\ $Z^{\Ja\Jb,\Jc}=0$,
and the gauge group generators~\eqref{XP} take the form
\bea
(X_{\Ja\Jb}){}_{\Jc}{}^{\Jd} &=&
\delta^{\Jd}_{[\Ja}\,Y^{\phantom{\Jd}}_{\Jb]\Jc} 
\;.
\label{CSO}
\eea
In this case, the quadratic constraint \eqref{quadratic} is automatically satisfied,
thus every symmetric matrix $Y_{\Ja\Jb}$ defines a viable gauging.
Fixing the ${\rm SL}(5)$ symmetry (and possibly rescaling the gauge coupling constant), 
this matrix can be brought into the form
\begin{align}
  Y_{\Ja\Jb} &= 
  \diag(\,\underbrace{1, \dots,}_{p}\underbrace{-1,\dots,}_{q} \underbrace{0, \dots}_{r}\,) \;,
  \label{YinCSO}
\end{align}
with $p+q+r=5$. The corresponding gauge group is
\begin{align}
   {\rm G}_0={\rm CSO}(p,q,r)={\rm SO}(p,q) \ltimes \mathbbm{R}^{(p+q)\cdot r} \; ,
\end{align}
where the abelian part combines $r$ vectors under ${\rm SO}(p,q)$. 
This completely classifies the gaugings in this sector.
The scalar potential~\eqref{LMP} reduces to
\bea
V&=& \ft1{64}
\Big(2{\cal M}^{\Ja\Jb}Y_{\Jb\Jc}{\cal M}^{\Jc\Jd}Y_{\Jd\Ja}-
({\cal M}^{\Ja\Jb}Y_{\Ja\Jb})^{2}\Big)\;.
\label{potCSO}
\eea
{}From Table~\ref{dof} one reads off the spectrum of these theories ($s=0$, $t=5-r$):
after gauge fixing
it consists of 10 vectors together with $r$ massless two-forms and
$5-r$ selfdual massive three-forms.
In particular, a nondegenerate $Y_{\Ja\Jb}$ ($r=0$)
corresponds to the semisimple gauge groups ${\rm SO}(5)$, 
${\rm SO}(4,1)$ and ${\rm SO}(3,2)$
that have originally been constructed 
exclusively in terms of vector and three-form 
fields~\cite{Pernici:1984xx,Pernici:1984zw}. 

The ${\rm SO}(5)$ gauged theory has a higher-dimensional interpretation
as reduction of $D=11$ supergravity on the sphere 
$S^{4}$~\cite{Pilch:1984xy,Nastase:1999cb,Nastase:1999kf}.
Accordingly, its potential~(\ref{potCSO})
admits a maximally supersymmetric AdS$_{7}$ ground state.
The theories with ${\rm CSO}(p,q,r)$ gauge groups
are related to the compactifications on the (noncompact) manifolds $H^{p,q} \circ T^r$
\cite{Hull:1988jw}.
These are the four-dimensional hypersurfaces of $\mathbb{R}^5$ defined by
\begin{align}
   Y_{\Ja\Jb} \, v^\Ja v^\Jb &= 1 \; , & &   v^\Ja \in \mathbb{R}^5 \; .
\end{align}
A particularly interesting example is the ${\rm CSO}(4,0,1)$ theory
which corresponds to the $S^{3}$ compactification of the ten-dimensional 
type IIA theory. The bosonic part of this theory has previously been 
constructed in~\cite{Cvetic:2000ah}. In order to derive its scalar potential
from~\eqref{potCSO} it is useful to parametrize the 
coset representative ${\cal V}$ as
\bea
{\cal V} &=& e^{b_{m}t^{m}}\,V_{4}\,e^{\phi\, t_{0}}\;, 
\label{M4}
\eea
where $V_{4}$ is an ${\rm SL}(4)/{\rm SO}(4)$ matrix and $t_{0}$, $t^{m}$
denote the ${\rm SO}(1,1)$ and four nilpotent generators, respectively,
in the decomposition ${\rm SL}(5)\rightarrow {\rm SL}(4)\times{\rm SO}(1,1)$. 
For the matrix ${\cal M}$ this yields a block decomposition into
\bea
{\cal M}_{\Ja\Jb} &=& \left(
\begin{array}{cc}
e^{-2\phi}\,M_{mn} + e^{8\phi}\, b_{m} b_{n} & e^{8\phi}\, b_{m}\\
e^{8\phi}\, b_{n} & e^{8\phi}
\end{array}
\right)
\eea
with $M=V^{{\rm \vphantom{T}}}_{4}\,V_{4}^{{\rm T}}$. 
Plugging this into~\eqref{potCSO} 
with $Y_{\Ja\Jb}={\rm diag}(1,1,1,1,0)$
yields the potential
\bea
V &=& \ft1{64}e^{4\phi}\,
\Big(2\,M^{mn}\delta_{nk}M^{kl}\delta_{lm}-(M^{mn}\delta_{mn})^{2}\Big)\;,
\label{potIIA}
\eea
(where $M_{mk}M^{kn}=\delta_{m}^{n}$)
in agreement with~\cite{Cvetic:2000ah}. The presence of the dilaton prefactor
$e^{4\phi}$ shows that this potential does not admit any stationary points,
rather the ground state of this theory is given by a domain wall solution
corresponding to the (warped) $S^{3}$ reduction of the 
type IIA theory~\cite{Cvetic:2000ah,Bergshoeff:2004nq}.

We can finally determine all the stationary points
of the scalar potentials~\eqref{potCSO} in this sector of gaugings.
The variation of the potential has been given in~\eqref{VaryV}.
Since $Z^{\Ja\Jb,\Jc}=0$, the tensors $C^{ab}$, $C^{[ab]}{}_{(cd)}$
vanish such that requiring $\delta_{\Sigma} V=0$ reduces to the matrix equation
\bea
2{\bf B}^{2}- B\, {\bf B} 
&=& \ft15\, {\rm Tr}(2{\bf B}^{2}- B\, {\bf B} )\,\mathbb{I}_{5}\;,
\label{matrix}
\eea
for the traceless symmetric matrix ${\bf B}=B^{[ab]}{}_{[cd]}$, where
$\mathbb{I}_{5}$ denotes the $5\times5$ unit matrix. 
According to~\eqref{BC} ${\bf B}$ is related by 
$\sqrt{2}{\bf Y}={\bf B}+B\,\mathbb{I}_{5}$ to the matrix ${\bf Y}=Y_{[\ja\jb],[\jc\jd]}$.
Fixing the local ${\rm USp}(4)$-invariance
the matrix ${\bf B}$ can be brought into diagonal form. Equation~\eqref{matrix}
then has only three inequivalent solutions
\bea
{\bf B}\propto {\rm diag}(0,0,0,0,0) &\Longrightarrow&
{\bf Y}= {\rm diag}(1,1,1,1,1) \;,
\nonumber\\
{\bf B}\propto {\rm diag}(1,1,1,1,-4)&\Longrightarrow&
{\bf Y}= 2^{-1/5}\,{\rm diag}(1,1,1,1,2) \;,
\nonumber\\
{\bf B}\propto {\rm diag}(1,1,1,-3/2,-3/2) &\Longrightarrow&
{\bf Y}= {\rm diag}(0,0,0,1,1) \;.
\label{statpoints}
\eea
The first two solutions correspond to the ${\rm SO}(5)$ and the ${\rm SO}(4)$
invariant stationary points of the theory with gauge group 
${\rm SO}(5)$~\cite{Pernici:1984xx,Pernici:1984zw}.
The third solution is a stationary point
in the ${\rm CSO}(2,0,3)$ gauged theory. We will come back to this 
in section~\ref{sec:YZ} and show that it
gives rise to a Minkowski vacuum related to a 
Scherk-Schwarz reduction from eight dimensions.

Analyzing the remaining supersymmetry of these vacua we note 
that in this sector of theories $A_{1}^{\ja\jb}\propto\Omega^{\ja\jb}$.
According to~(\ref{susy_groundstate2}) thus supersymmetry is
either completely preserved (${\cal N}=4$) or completely broken 
(${\cal N}=0$). Only the first stationary point in~\eqref{statpoints}
preserves all supersymmetries: this is the maximally supersymmetric
AdS$_{7}$ vacuum mentioned above.

\subsection{Gaugings in the ${\bf \overline{40}}$ representation:~ \\ 
${\rm SO}(p,4\!-\!p)$ and
${\rm CSO}(p,q,4\!-\!p\!-\!q)$}
\label{SecExNoY}

Another sector of gaugings is characterized by restricting
the embedding tensor to the ${\bf \overline{40}}$ representation
of ${\rm SL}(5)$, i.e.\ setting $Y_{\Ja\Jb}=0$. 
These gaugings are parametrized by a tensor $Z^{\Ja\Jb,\Jc}$
for which the quadratic constraint~\eqref{quadratic} reduces to
\bea
\epsilon_{\Ja\Je\Jf\Jg\Jh}\,Z^{\Je\Jf,\Jb}Z^{\Jg\Jh,\Jc} &=& 0 \;.
\label{qZ}
\eea
Rather than attempting a 
complete classification of these theories we will present a representative
class of examples.
Specifically, we consider gaugings with the tensor $Z^{\Ja\Jb,\Jc}$ given by
\begin{align}
   Z^{\Ja\Jb,\Jc} \, &= \, v^{[\Ja} \, w^{\Jb]\Jc}
   \;,
   \label{ZeqVW}
\end{align}
in terms of a vector $v^\Ja$ and a symmetric matrix $w^{\Ja\Jb}=w^{(\Ja\Jb)}$. 
This ansatz automatically solves the quadratic constraint~\eqref{qZ} and thus
defines a class of viable gaugings.
The ${\rm SL}(5)$ symmetry can be used to further bring $v^{\Ja}$ into the form 
$v^{\Ja}=\delta^{\Ja}_{5}$ introducing the index split 
${\Ja}=({i},{5})$, $i=1, \dots, 4$.
The remaining ${\rm SL}(4)$ freedom can be fixed
by diagonalizing the corresponding $4\times4$ block $w^{ij}$
\begin{align}
  w^{ij} &= 
  \diag(\,\underbrace{1, \dots,}_{p}\underbrace{-1,\dots,}_{q} \underbrace{0, \dots}_{r}\,) \; .
\end{align}
For simplicity we restrict to cases with $w^{i5}=w^{55}=0$.
The gauge group generators then take the form
\bea
(X_{ij}){}_{k}{}^{l} &=&
2\epsilon_{ijkm}\, w^{ml}
\;,
\label{cso4}
\eea
and generate the group~${\rm CSO}(p,q,r)$ with $p+q+r=4$.
According to Table~\ref{dof}, these theories contain only
vector and two-forms, $4\!-\!r$ of which become massive after gauge 
fixing. The scalar potential is obtained from~\eqref{LMP}
and in the parametrization of~\eqref{M4} takes the form
\bea
V &=& \ft14e^{14\phi}\,b_{m}w^{mk}M_{kl}\,w^{ln}\,b_{n}
+\ft14e^{4\phi}\,\Big(2\,M_{mn}w^{nk}M_{kl}w^{lm}-(M_{mn}w^{mn})^{2}\Big)\;.
\label{potIIB}
\eea
A particularly interesting case  is the theory with $r=0$ and
compact gauge group ${\rm SO}(4)$.
The existence of this maximal supergravity in seven dimensions was 
anticipated already in \cite{Boonstra:1998mp} in the context of holography to 
six-dimensional super Yang-Mills theory. 
Indeed, its spectrum should consist of vector and two-form 
tensor fields only (cf.\ Table~IV in~\cite{Morales:2004xc}).
Its higher-dimensional origin is a (warped) $S^3$ reduction of type IIB supergravity.
Again, this is consistent with the fact that 
due to the presence of the dilaton prefactor
the potential~\eqref{potIIB} in this case
does not admit any stationary points 
but only a domain wall solution.
So far, only the ${\cal N}=2$ truncation of this theory had been 
constructed~\cite{Salam:1983fa,Cvetic:2000dm}, in which the scalar manifold
truncates to an ${\rm GL}(4)/{\rm SO}(4)$ coset space and only a single (massless)
two-form is retained in the spectrum.

In analogy to the discussion of the last section it seems natural that the 
other ${\rm CSO}(p,q,r)$ gaugings in this sector
are related to reductions of the type IIB theory over the 
noncompact manifolds $H^{p,q} \circ T^r$. In particular, the 
potential~\eqref{potIIB} of the ${\rm CSO}(2,0,2)$ theory admits a 
stationary point with vanishing potential.
This is related to the Minkowski vacuum obtained by
Scherk-Schwarz reduction from eight dimensions as we will discuss in the next section.

\subsection{Further examples}
\label{sec:YZ}

We will finally indicate a more systematic approach towards 
classifying the general gaugings with an embedding tensor combining
parts in the ${\bf 15}$ and the ${\bf \overline{40}}$ representation.
To this end, we go to the special basis introduced in section~\ref{sec:gfixing},
in which the only nonvanishing components of the embedding tensor are given by
\bea
Y_{xy}\;,\qquad Z^{x(\alpha,\beta)}\;,\qquad Z^{\alpha\beta,\gamma}
\;,
\label{nonv}
\eea
with ${\rm rank}\, Y\equiv t$, and the range of indices  $x, y=1, \dots, t$ and 
$\alpha, \beta=t\!+\!1, \dots, 5$ . 
Further fixing (part of) the global ${\rm SL}(5)$ symmetry, 
the tensor $Y_{xy}$ can always be brought into the standard form 
\begin{align}
  Y_{xy} &= 
  \diag(\,\underbrace{1, \dots,}_{p}\underbrace{-1,\dots}_{q}\,) \;.
  \label{Yxy}
\end{align}
The possible gaugings can then systematically be found by 
scanning the different values of $t$, $p$, and $q$,
and determining the real 
solutions of the quadratic constraint~\eqref{quadraticZ}.
We will in the following discuss a (representative rather than complete) 
number of examples for the different values of $t$. 
A list of our findings is collected in Table~\ref{TabSumEx}.

\mathversion{bold}
\subsection*{$t=5$}
\mathversion{normal}

{}From~\eqref{nonv} one reads off that a nondegenerate matrix $Y_{\Ja\Jb}$ implies
a vanishing tensor $Z^{\Ja\Jb,\Jc}$. 
Thus we are back to the situation discussed 
in section~\ref{SecExNoZ}. The possible gauge groups are 
${\rm SO}(5)$,  ${\rm SO}(4,1)$, and  ${\rm SO}(3,2)$.

\mathversion{bold}
\subsection*{$t=4$}
\mathversion{normal}

{}The quadratic constraint~\eqref{quadraticZ} implies that also in this case 
the tensor $Z^{\Ja\Jb,\Jc}$ entirely vanishes. These gaugings
are again completely covered by the discussion of section~\ref{SecExNoZ}, 
with possible gauge groups 
${\rm CSO}(4,0,1)$,  ${\rm CSO}(3,1,1)$, and  ${\rm CSO}(2,2,1)$.

\mathversion{bold}
\subsection*{$t=3$}
\mathversion{normal}

Now we consider the cases $Y_{\Ja\Jb}=\diag(1,1,\pm 1,0,0)$. 
In this case the tensor $Z$ may have nonvanishing components
for which the quadratic constraint~\eqref{quadraticZ} 
imposes
\bea
\epsilon_{xyz}\,Z^{y\alpha,\gamma}\,\epsilon_{\gamma\delta}\,Z^{z\delta,\beta}
&=& \ft18\,Y_{xu}Z^{u\alpha,\beta} 
\;.
\label{quadraticZ23}
\eea
For $Z=0$, these gaugings have been discussed in section~\ref{SecExNoZ}, 
with possible gauge groups ${\rm CSO}(3,0,2)$ and  ${\rm CSO}(2,1,2)$.
There, gauge group generators take the form
\bea
L_\Ja{}^\Jb  &=&
\left( \begin{array}{cc}  \lambda^{z} (t^{z})_{x}{}^{y} & Q_{x\alpha} \\
0_{2 \times 3} & 0_{2\times 2} \end{array} \right) \;,\qquad
\lambda^{z}\in\mathbb{R}\;,\quad Q_{x\alpha}\in\mathbb{R}\;,
   \label{cso302}      
\eea
where $(t^{z})_{x}{}^{y}=\epsilon^{zyu}Y_{ux}$ 
denote the generators of the adjoint representation 
of the semisimple part $\mathfrak{so}(p,3\!-\!p)$ and 
the $Q_{x\alpha}$ parametrize the 6 nilpotent
generators transforming as a couple of ${\bf 3}$ vectors 
under $\mathfrak{so}(p,3\!-\!p)$.
The components~$Z^{\ka\kb,\kc}$ are not constrained by~\eqref{quadraticZ23}
and may be set to arbitrary values $Z^{\ka\kb,\kc} = \epsilon^{\ka\kb} v^\kc$ 
parametrized by a two-component vector~$v^{\ka}$
without altering the form~\eqref{cso302} of the gauge group.
For the remaining components $Z^{\la\alpha,\beta}$, 
equation~\eqref{quadraticZ23} 
shows that the $2\times2$ matrices 
$(\Sigma^{\la}){}_\alpha{}^\beta \equiv -16 \epsilon_{\alpha\gamma}Z^{\la\gamma,\beta}$
satisfy the algebra
\bea
   [\Sigma^x,\Sigma^y] &= 2 \epsilon^{xyu}Y_{uz}\, \Sigma^z \; , 
\eea
i.e.\ yield a representation of the algebra $\mathfrak{so}(3)$ or 
$\mathfrak{so}(2,1)$, respectively, depending on the signature of $Y_{uz}$.
A real nonvanishing solution of~\eqref{quadraticZ23} thus can only exist
in the $\mathfrak{so}(2,1)$ sector, i.e.\
for $Y_{\Ja\Jb}=\diag(1,1,-1,0,0)$. It is given by 
$Z^{\la\ka,\kb} = - \, \frac 1 {16} \, \epsilon^{\ka\kc} \, (\Sigma^{\la})_\kc{}^\kb$
with the $\Sigma^{x}$ expressed in terms of the Pauli matrices as
\begin{align}
   \Sigma^1 &= \sigma_{1} \; ,\quad
   \Sigma^2 = \sigma_{3}  \; ,\quad
   \Sigma^3 = i\sigma_{2}  \;,
\end{align}
and providing a real representation of $\mathfrak{so}(2,1)$.
In this case, the gauge group generators 
schematically take the form
\bea
{{L_\Ja}}^\Jb  &= &
\left( \begin{array}{cc} \lambda^{z} (t^{z})_{x}{}^{y} & Q^{(4)}_{\vphantom{[]}x\alpha} \\
0_{2 \times 3} & \ft12\lambda^{z}\,(\Sigma^{z})_{\alpha}{}^{\beta} \end{array} \right) \;,
   \label{notcso302}      
\eea
such that the semisimple part $\mathfrak{so}(2,1)$ is 
embedded into the diagonal. The nilpotent generators $Q_{x\alpha}$
now transform in the tensor product ${\bf 3}\otimes{\bf 2}={\bf 2}+{\bf 4}$ 
of $\mathfrak{so}(2,1)$ and moreover
turn out to be projected onto the irreducible ${\bf 4}$ representation. 
Compared to~\eqref{cso302},
the gauge group thus shrinks to
\bea
\mathfrak{so}(2,1) \ltimes \mathbb{R}^{4}\;.
\eea
Again, further switching on $Z^{\ka\kb,\kc}$ does not 
change the form of the algebra.
None of the theories in this sector possesses a stationary 
point in its scalar potential.

\mathversion{bold}
\subsection*{$t=2$}
\mathversion{normal}

In the case $Y_{\Ja\Jb}=(1,\pm 1,0,0,0)$ only
the $Z^{\ka\kb,\kc}$ components are allowed to be nonzero 
in order to fulfill the quadratic constraint \eqref{quadraticZ}.
These components can be parametrized by a traceless 
matrix ${Z_{\ka}}^{\kb}$ as
\begin{align}
   Z^{\ka\kb,\kc} &= \ft18\epsilon^{\ka\kb\kd} {Z_{\kd}}^\kb \;.
\end{align}
For this solution the gauge generators take the form
\begin{align}
    {{L_\Ja}}^\Jb  &= \left( \begin{array}{cc}  \lambda \, t_{2\,x}{}^{y} & Q_{x}{}^{\alpha} \\
0_{3 \times 2} &  \lambda \, Z_{\alpha}{}^{\beta} \end{array} \right) \;,
\qquad
\lambda\in\mathbb{R}\;,\quad Q_{x}{}^{\alpha}\in\mathbb{R}\;,
   \label{GenYZr2}
\end{align}
where $t_{2}=\Big(\!\!\!{\footnotesize\begin{array}{rc} 0 & \!\!\!\!\!1 
   \\[-1ex] \mp 1 & \!\!\!\!\!0 \end{array} }\!\!\Big)$ denotes
a generator of $\mathfrak{so}(2)$ or  $\mathfrak{so}(1,1)$, respectively,
and $Q_{x}{}^{\alpha}$ parametrizes  a generically unconstrained block  of six translations.
Thus, generically the gauge group ${\rm G}_0$ in this case is seven-dimensional, 
namely either ${\rm G}_0={\rm SO}(2) \ltimes \mathbb{R}^6$ or ${\rm G}_0={\rm SO}(1,1) \ltimes \mathbb{R}^6$.
The number of independent translations is 
reduced in case the equation
\bea
t_{2}\, Q - Q Z &=&0  \;,
\label{redu}
\eea
has nontrivial solutions $Q$. In this case, the gauge group shrinks to 
${\rm G}_0={\rm SO}(2) \ltimes \mathbb{R}^s$ or ${\rm G}_0={\rm SO}(1,1) \ltimes \mathbb{R}^s$, 
with $s=4,5$.
The scalar potential in this sector can be computed from~\eqref{LMP}
and takes the form
\bea
V &=& \ft1{64}\,
\Big( 2 {\rm Tr}\,[{\hat Y}{}^{2}] - ({\rm Tr}\,\hat Y)^{2} + 
2 \,({\rm det}{\cal M}_{\alpha\beta})\, {\rm Tr}\,[\hat Z^{2}] \Big)
\;,
\label{potSS}
\eea
in terms of the 
matrices $\hat Y_{x}{}^{y}=Y_{xz}{\cal M}^{zy}$
and 
$\hat Z_{\alpha}{}^{\beta}=
Z_{(\alpha}{}^{\gamma}{\cal M}_{\delta)\gamma}{\cal M}^{\delta\beta}$.
Here, ${\cal M}^{xy}$ and ${\cal M}_{\alpha\beta}$ denote the
diagonal blocks of the symmetric unimodular matrix 
defined in~\eqref{DefCalM}, and 
${\cal M}_{\alpha\gamma}{\cal M}^{\gamma\beta}=\delta_{\alpha}^{\beta}$.
Since the matrix  $\hat Y_{x}{}^{y}$ has only two nonvanishing eigenvalues,
this potential is positive definite. In particular, this
implies that $V=0$ is a sufficient condition for a stationary point.
It further follows from~\eqref{potSS} that $V$ only vanishes
for  $\hat Y_{x}{}^{y}\propto \delta_{x}{}^{y}$ and 
$Z_{(\alpha}{}^{\gamma}{\cal M}_{\delta)\gamma}=0$, i.e.\ for
compact choice of $t_{2}$ and $Z$. 
With vanishing $Z$ or vanishing $t_{2}$ one recovers the
Minkowski vacua in the ${\rm CSO}(2,0,3)$ and the ${\rm CSO}(2,0,2)$
theory, respectively, discussed in 
sections~\ref{SecExNoZ} and \ref{SecExNoY} above.

In turn, every compact choice of $t_{2}$ and $Z$ defines a theory with 
a Minkowski vacuum in the potential. The gravitino masses and thereby the 
remaining supersymmetries at this ground state are determined from 
the eigenvalues of $A_{1\,ab}$~\eqref{susy_groundstate2}
according to
\bea
m_{\pm}^{2} &=&   
\frac1{1600}\,\bigg(1\pm \sqrt{-\frac12{\rm Tr}Z^{2}}\;\bigg)^{2} \;.
\eea
Half of the supersymmetry (${\cal N}=2$) is thus preserved iff ${\rm Tr}Z^{2}=-2$.
With~\eqref{redu} one finds that precisely at this value the rank of the 
gauge group decreases from 7 down to~5; the group then is ${\rm CSO}(2,0,2)$.

All the gaugings in this sector have a well defined higher-dimensional origin,
namely they descend by Scherk-Schwarz reduction~\cite{Scherk:1979zr} 
from the maximal theory in eight dimensions. Indeed, Scherk-Schwarz reduction 
singles out one generator from the ${\rm SL}(2)\times {\rm SL}(3)$
global symmetry group of the eight-dimensional theory~\cite{Salam:1984ft}. 
With the seven-dimensional embedding tensor branching as
\bea
Y:\quad{\bf 15}&\rightarrow& ({\bf 3},{\bf 1})+({\bf 2},{\bf 3})+({\bf 1},{\bf 6}) \;,
\nonumber\\
Z:\quad{\bf \overline{40}}&\rightarrow& ({\bf 1},{\bf \overline{3}})+({\bf 1},{\bf 8})+
({\bf 2},{\bf 1}) +
({\bf 2},{\bf 3})+ ({\bf 2},{\bf \overline{6}})+({\bf 3},{\bf \overline{3}})\;,
\eea
a Scherk-Schwarz gauging corresponds to
switching on components $({\bf 3},{\bf 1})+({\bf 1},{\bf 8})$
in the adjoint representation of~${\rm SL}(2)\times {\rm SL}(3)$.
This precisely amounts to the parametrization in terms of 
matrices $Y_{xy}$, $Z_{\alpha}{}^{\beta}$
introduced above. 
We have seen that for compact choice of $t_{2}$ and $Z$,
the potential~\eqref{potSS} admits a Minkowski ground state
as expected from the Scherk-Schwarz origin.
Moreover, we have shown that for a particular ratio between
the norms of $t_{2}$ and $Z$, this ground state preserves $1/2$
of the supersymmetries.

\mathversion{bold}
\subsection*{$t=1,0$}
\mathversion{normal}
As $t$ becomes smaller, the consequences of the quadratic constraint~\eqref{quadraticZ}
become more involved. We refrain from attempting a complete classification in this sector
and refer to the examples that we have discussed 
in sections~\ref{SecExNoZ} and \ref{SecExNoY} above.

\begin{table}[tb]
   \begin{center}
      \begin{tabular}{c|c|c|c|c|c|c}
      $t$ & $Y_{\Ja\Jb}$ &  $Z^{\ka\kb,\kc}$  & $Z^{x\ka,\kb}$  &\quad gauge group 
      & stat.\ point & susy \\[0.1cm]
\hline
 $5$&$(+\!+\!+\!+\!+)$ &  &  & ${\rm SO}(5)$ & $\times$\,,\,$\times$ & 4\,,\,0 \\[0.2cm]
 $5$&$(+\!+\!+\!+\!-)$ &  &  & ${\rm SO}(4,1)$ & $-$ &  \\[0.2cm]
 $5$&$(+\!+\!+\!-\!-)$ &  &  & ${\rm SO}(3,2)$ & $-$ &  \\[0.2cm]
 \hline
 $4$&$(+\!+\!+\!+\,0)$ &  & & ${\rm CSO}(4,0,1)$ & $-$ &  \\[0.2cm]
 $4$&$(+\!+\!+\!-\,0)$ &  &  & ${\rm CSO}(3,1,1)$ & $-$ &  \\[0.2cm]
 $4$&$(+\!+\!-\!-\,0)$ &  &  & ${\rm CSO}(2,2,1)$ & $-$ &  \\[0.2cm]
\hline
 $3$&$(+\!+\!+\,0\,0)$ &$\epsilon^{\ka\kb}v^{\kc}$ & & ${\rm CSO}(3,0,2)$ & $-$ &  \\[0.2cm]
 $3$&$(+\!+\!-\,0\,0)$ & $\epsilon^{\ka\kb}v^{\kc}$ &  & ${\rm CSO}(2,1,2)$ & $-$ &  \\[0.2cm]
 $3$&$(+\!+\!-\,0\,0)$ & $\epsilon^{\ka\kb}v^{\kc}$ &
\!\! $\frac 1{16}\epsilon^{\kc\ka}(\Sigma^{\la})_\kc{}^\kb$\!
  & ${\rm SO}(2,1)\!\ltimes\! \mathbb{R}^4$ & $-$ &  \\[0.2cm]
 \hline
 $2$&$(+\!+\,0\,0\,0)$ & $\ft18\epsilon^{\ka\kb\kd} {Z_{\kd}}^\kc$ &
  & ${\rm SO}(2) \!\ltimes\! \mathbb{R}^s$ & $\times$ &
 $2\rightarrow0$  \\[0.2cm]
 $2$&$(+\!-\,0\,0\,0)$ & $\ft18\epsilon^{\ka\kb\kd} {Z_{\kd}}^\kc$ &
   & ${\rm SO}(1,1) \!\ltimes\! \mathbb{R}^s$ & $-$ &  \\[0.2cm]
\hline
 $1$&$(+\,0\,0\,0\,0)$ &&& ${\rm CSO}(1,0,4)$ & $-$ &  \\[0.2cm]
\hline
       $0$ &$(0\,0\,0\,0\,0)$  &  $v^{[\alpha} \, w^{\beta]\gamma}$ &
     & ${\rm SO}(p,4\!-\!p)$ 
       & $-$      & \\[0.2cm] 
       $0$ &$(0\,0\,0\,0\,0)$  &  $v^{[\alpha} \, w^{\beta]\gamma}$ &
       & $\begin{array}{c}
       {\rm CSO}(p,q,r)\\[-.5ex]
       (p\!+\!q\!+\!r=4) \end{array}$ 
       & $\begin{array}{c} \times\\[-.5ex](p\!=\!2\!=\!r) \end{array}$      
        & 0
      \end{tabular}
   \end{center}
   \caption{\label{TabSumEx}{\small Examples for gaugings of $D=7$ maximal supergravity.}}
\end{table}

\section{Conclusions}
\setcounter{equation}{0}

In this paper we have presented the possible deformations
of maximal seven-dimensional supergravity. They are described
by a universal Lagrangian~\eqref{L} 
that combines vector, two-form and three-form tensor fields
transforming in the ${\bf\overline{10}}$, ${\bf{5}}$, and
${\bf\overline{5}}$ representation of ${\rm SL}(5)$, respectively.
The Lagrangian is invariant under an extended set
of nonabelian gauge transformations as well as under maximal 
supersymmetry.

The gaugings are entirely parametrized in terms of an embedding
tensor $\Theta$ which describes the embedding of the gauge group
${\rm G}_{0}$ into ${\rm SL}(5)$. At the same time, its irreducible components
in the ${\bf\overline{40}}$ and ${\bf{15}}$ representation of ${\rm SL}(5)$
induce St\"uckelberg type couplings between two-form fields and 
the vector fields and between the three-form and the two-form fields, 
respectively.
Altogether this gives rise to an extended 
vector/tensor system subject to a set of nonabelian
gauge transformations~\eqref{gauge} which ensures that the total number
of degrees of freedom is independent of the specific form of the
embedding tensor as required by supersymmetry.
Upon choosing a specific $\Theta$ and possibly fixing
part of the tensor gauge symmetry, the degrees of freedom are
properly distributed among the different forms.
This universal formulation thus accommodates theories 
with seemingly rather different field content.
This completes the seven-dimensional picture 
which neatly fits the pattern realized in other space-time 
dimensions~\cite{Nicolai:2000sc,deWit:2002vt,deWit:2004nw,deWit:2005hv}.

As particular examples we have recovered 
in this framework the known seven-dimensional 
gaugings as well as a number of 
new examples. Some of these theories have a definite
higher-dimensional origin, such as the Scherk-Schwarz 
and sphere compactifications.
In eight space-time dimensions, the possible compactifications
of $D=11$ supergravity on three-manifolds have been
analyzed in~\cite{Bergshoeff:2003ri} and matched
with the corresponding gauged supergravities.
It would be very interesting to extend this analysis
to the seven-dimensional case, in particular providing
a higher-dimensional origin for all the theories collected in 
Table~\ref{TabSumEx}.
More ambitiously, one may aim at understanding the role 
of the full embedding tensor $\Theta$ which parametrizes the 
different seven-dimensional gaugings directly in the 
eleven-dimensional theory. For the four-dimensional
gaugings in which $\Theta$ generically transforms
in the ${\bf 912}$ representation of the duality group ${\rm E}_{7(7)}$
this has been achieved in a few sectors~\cite{deWit:2003hq,DAuria:2005er} 
where particular 
components of $\Theta$ have been identified with internal fluxes and twists. 
Extending this correspondence to the full representation of the 
embedding tensor might in particular 
elucidate the possible role of the duality 
groups ${\rm E}_{d(d)}$ in eleven dimensions.

\bigskip

\noindent
{\bf Acknowledgement}\\
\noindent
We thank B. de Wit and M. Trigiante for discussions.
This work is partly supported by the EU contracts 
MRTN-CT-2004-503369 and
MRTN-CT-2004-512194,
and the DFG grant SA 1336/1-1. 

\bigskip
\bigskip

\begin{appendix}

\section*{Appendix}
\section{General vector/tensor gauge transformations}
\setcounter{equation}{0}
\label{APP-modified}

In this appendix we briefly summarize the
results of~\cite{deWit:2005hv} 
on the general form of vector/tensor gauge transformations 
in arbitrary space-time dimensions
and translate them into a more convenient basis.
Generically vector fields
and two-form tensor fields transform in 
different representations of the symmetry group ${\rm G}$
of the ungauged theory.
In this appendix, we label the vector representation $A_{\mu}{}^{\hat M}$
by indices $\hat M, \hat N, \dots$, the tensor representation $B_{\mu\nu\,I}$
by indices $I, J, \dots$, and the adjoint representation of ${\rm G}$ by
indices $\alpha, \beta, \dots$. In particular, the embedding tensor~(\ref{X-theta})
characterizing the gauging in general has the index
structure $\Theta_{\hat M}{}^{\alpha}$.

The two-form tensor fields $B_{\mu\nu\,I}$ generically transform
in an irreducible component of the
symmetric tensor product of two vector representations
with an explicit relation
\bea
X_{(\hat M\hat N)}{}^{\hat P} &=& d_{I,\hat M\hat N}\,Z^{\hat P,I}
\;.
\label{A1}
\eea
Here, 
$X_{\hat M\hat N}{}^{\hat P}\equiv
\Theta_{\hat M}{}^{\alpha}\,t_{\alpha\,\hat N}{}^{\hat P}$
generalizes (\ref{X-theta}) 
with the $\mathfrak{g}={\rm Lie}\, {\rm G}$ generators $t_{\alpha}$,
and encodes the ``structure constants'' of the gauged theory
while $d_{I,\hat M\hat N}$ is a ${\rm G}$-invariant tensor
projecting the symmetric vector product onto the tensor field 
representation.
The tensor $Z^{\hat M,I}$ represents a component of the embedding tensor
and equation~\eqref{A1} can be taken as its definition.
In seven dimensions, this equation
takes the specific form~(\ref{Csym}) discussed in the main text,
with the $d$-symbol provided by the $\epsilon$-tensor.
The three-form fields generically appear 
with index structure $S_{\mu\nu\rho\,I}{}^{\hat M}$, i.e.\ they
take values in the tensor product of vector and two-form tensor 
representations. More specifically, they
appear under projection $Y_{I,\hat M}{}^J\,S_{J}{}^{\hat M}$
with the tensor 
\begin{equation}
\label{eq:Y-def}
Y_{I,\hat M}{}^J \equiv X_{\hat MI}{}^J + 2\,d_{I,\hat M\hat N}\,Z^{\hat N,J}\,.
\end{equation}
In the maximal seven-dimensional theory,
with vector fields transforming in the ${\bf \overline{10}}$ and 
two-form tensors in the ${\bf 5}$, 
equation~\eqref{idY} states that
the tensor $Y_{I,\hat M}{}^J$ reduces to
\bea
Y_{K,[MN]}{}^{L} &=&
\delta^{L}_{[M}\,Y^{\phantom{L}}_{N]K} \;,
\eea
in terms of the component $Y_{MN}$ of the embedding tensor~(\ref{linear}).
This implies that the three-form tensors $S_{\mu\nu\rho\,L}{}^{MN}$ 
always appear projected according to
$Y_{K,[MN]}{}^{L}\,S_{L}{}^{MN}
=Y_{NK}\,S_{M}{}^{MN} \equiv Y_{NK}\,S^{N}$ 
and reflects the fact that the three-form fields in seven dimensions 
transform in the representation ${\bf \overline 5}$
dual to the two-form tensor fields.

Generic vector and tensor gauge transformations
are most conveniently described in terms
of the ``covariant variations''
introduced in~(\ref{covV}) in the main text
\bea
\Delta A_\mu{}^{\hat M}
&\equiv& \delta A_\mu{}^{\hat M} 
\;,
\nonumber\\[1ex]
\Delta B_{\mu\nu\,I} &\equiv&
\delta B_{\mu\nu\,I}
-2\, d_{I,\hat P\hat Q}\,A_{[\mu}{}^{\hat P}\,\delta A_{\nu]}{}^{\hat Q}
\;,
\nonumber\\[1ex]
\Delta S_{\mu\nu\rho\,I}{}^{\hat M} &\equiv&
\delta S_{\mu\nu\rho\,I}{}^{\hat M} 
-3\,B_{[\mu\nu\,I} \,\delta A_{\rho]}{}^{{\hat M}}
- 2  d_{I,\hat P\hat Q}
A_{[\mu}{}^{\hat M} A_{\nu}{}^{\hat P}\,\delta A_{\rho]}{}^{\hat Q}
\;,
\label{D1}
\eea
and given by
\bea
\Delta A_\mu{}^{\hat M} &=&
D_\mu\Lambda^{\hat M}
- g\,Z^{\hat M,I}\,\Xi_{\mu\,I}
\;,
\nonumber\\[1ex]
\Delta B_{\mu\nu\,I} 
&=&
2\,D_{[\mu} \Xi_{\nu]I} 
-2d_{I,\hat P\hat Q}\, \Lambda^{\hat P}\,{\cal H}^{(2)}_{\mu\nu}{}^{\hat Q}
- g\, Y_{I,{\hat M}}{}^J \,\Phi_{\mu\nu\,J}{}^{\hat M}
\;,
\nonumber\\[1ex]
\Delta S_{\mu\nu\rho\,I} {}^{{\hat M}}
&=&
3\,D_{[\mu}\Phi_{\nu\rho]\,I}{}^{\hat M} 
 + 3\,{\cal H}^{(2)}_{[\mu\nu}{}^{\hat M}\,\Xi_{\rho]I} 
+ \Lambda^{\hat M}\,{\cal H}^{(3)}_{\mu\nu\rho\,I}
\;,
\label{D2}
\eea
with gauge parameters $\Lambda^{\hat M}$, $\Xi_{\mu I}$, $\Phi_{\mu\nu\,I}{}^{\hat M}$,
and the covariant field strengths
\bea
{\cal H}^{(2)}_{\mu\nu}{}^{\hat M} &=& 
2\,\partial_{[\mu} A_{\nu]}{}^{\hat M} 
+ g\,X_{[\hat N\hat P]}{}^{\hat M} \,A_\mu{}^{\hat N} A_\nu{}^{\hat P}+ g\, 
Z^{{\hat M},I} \,B_{\mu\nu\,I}
\;,
\nonumber\\[2ex]
{\cal H}^{(3)}_{\mu\nu\rho\,I} &\equiv&
3\, D_{[\mu} B_{\nu\rho]\,I} +6
\,d_{I,\hat M\hat N}\,A_{[\mu}{}^{\hat M}(\pa_{\nu} A_{\rho]}{}^{\hat N}+ 
\ft13 g X_{[\hat P\hat Q]}{}^{\hat N}A_{\nu}{}^{\hat P}A_{\rho]}{}^{\hat Q})\nn\\
&&{} 
+ g\,Y_{I,\hat M}{}^J\, S_{\mu\nu\rho\,J}{}^{\hat M} \;,
\label{covAH}
\eea
The formulas of~\cite{deWit:2005hv} 
are recovered from~(\ref{D1}), (\ref{D2})
by modifying the gauge parameters
$\Xi_{\mu\,I}$, $ \Phi_{\mu\nu\,I}{}^{\hat M}$ according to
\bea
\Xi_{\mu\,I}&\rightarrow&
\Xi_{\mu\,I}-d_{I,\hat P\hat Q}\,\Lambda^{\hat P}\,A_\mu{}^{\hat Q}
\;,
\nonumber\\[1ex]
 \Phi_{\mu\nu\,I}{}^{\hat M}
 &\rightarrow&
\Phi_{\mu\nu\,I}{}^{\hat M}
 + A_{[\mu}{}^{\hat M}\,\Xi_{\nu]I} 
+ \Lambda^{\hat M}\,B_{\mu\nu\,I}
- \ft13\,d_{I,\hat P\hat Q}\,\Lambda^{\hat P}\,A_{[\mu}{}^{\hat M} A_{\nu]}{}^{\hat Q}
\;.
\eea
The covariant variations~(\ref{D1}) appear naturally 
in the variation of the covariant field strengths~(\ref{covAH})
as
\bea
\delta\,{\cal H}^{(2)}_{\mu\nu}{}^{\hat M} &=&
 2\,D_{[\mu}\, (\Delta A_{\nu]}{}^{\hat M}) 
 +g\, Z^{{\hat M},I} \,\Delta B_{\mu\nu\,I}
 \;,
 \nonumber\\[3ex]
 \delta\,{\cal H}^{(3)}_{\mu\nu\rho\,I} &=&
 3\, D_{[\mu} (\Delta B_{\nu\rho]\,I})
+ 6\,d_{I,\hat M\hat N}\,{\cal H}^{(2)}_{[\mu\nu}{}^{\hat M}\, \Delta A_{\rho]}{}^{\hat N}
+ g\,Y_{I\hat M}{}^J\, 
\Delta S_{\mu\nu\rho\,J}{}^{\hat M} \;.
\eea
Similarly, we have shown in the main text
that the variation of the topological term~(\ref{varCS}) 
also comes in terms of the covariant transformations~(\ref{D1}).
The advantage of formulating the gauge transformations~(\ref{D2}) 
in the new basis thus is that they 
keep the variation of the Lagrangian manifestly covariant.

\section{${\rm USp}(4)$ invariant tensors}
\setcounter{equation}{0}
\label{app:tensors}

We label the fundamental representation of
${\rm USp}(4)$ by indices $\ja$, $\jb$, \ldots \ running from $1$ to $4$.
The lowest bosonic representations of
${\rm USp}(4)$ have been 
collected in~\eqref{USp4Reps} built in terms of the fundamental representation.
In particular, 
the ${\bf 5}$ representation is 
given by an antisymmetric symplectic traceless tensor ${V_{{\bf 5}}}^{[\ja\jb]}$,
objects in the ${\bf 10}$ are described by a symmetric tensor ${V_{{\bf 10}}}^{(\ja\jb)}$, etc.

In this appendix we introduce a number of ${\rm USp}(4)$ invariant tensor 
which explicitly describe the projection of ${\rm USp}(4)$ tensor products
onto their irreducible components and derive some relations between them.
All of these tensors are constructed from the invariant 
symplectic form $\Omega_{\ja\jb}$ and the relations that they satisfy
can be straightforwardly derived form the properties of $\Omega_{\ja\jb}$. 
We have used these tensors extensively in the course of our calculations, 
while the final results in the main text are formulated 
explicitly in terms of $\Omega_{\ja\jb}$.

On the ${\bf 5}$ and ${\bf 10}$ representation 
of ${\rm USp}(4)$ there are nondegenerate symmetric forms given by
\begin{align}
   \delta_{[\ja\jb][\jc\jd]} &= - \Omega_{\ja [\jc} \Omega_{\jd] \jb} - \frac 1 4 \Omega_{\ja\jb} \Omega_{\jc\jd}  \; , &
 \nonumber\\
   \delta^{[\ja\jb][\jc\jd]} &= \left( \delta_{[\ja\jb][\jc\jd]} \right)^* = - \Omega^{\ja [\jc} \Omega^{\jd] \jb}  
                                                                         - \frac 1 4 \Omega^{\ja\jb} \Omega^{\jc\jd} \; , &
 \nonumber\\
   \delta^{[\ja\jb]}_{[\jc\jd]} &= 
   \delta^{[\ja}_{[\jc} \delta^{\jb]}_{\jd]} - \frac 1 4 \Omega_{\jc\jd} \Omega^{\ja\jb} 
   = 
   \delta^{\ja\jb}_{\jc\jd}  - \frac 1 4 \Omega_{\jc\jd} \Omega^{\ja\jb} \; , &
\nonumber\\[3ex] 
   \delta^{(\ja\jb)(\jc\jd)} &= \left( \delta_{(\ja\jb)(\jc\jd)} \right)^* = - \Omega^{\ja (\jc} \Omega^{\jd) \jb}  \; , \nonumber \\
    \delta_{(\ja\jb)(\jc\jd)} &= - \Omega_{\ja (\jc} \Omega_{\jd) \jb} \; , \nonumber \\[.8ex]
    \delta^{(\ja\jb)}_{(\jc\jd)} &= \delta^{(\ja}_{(\jc} \delta^{\jb)}_{\jd)} \; .
\end{align}
Note that $\delta^{[\ja\jb][\jc\jd]}$ is the inverse of $\delta_{[\ja\jb][\jc\jd]}$, i.e.
\begin{align}
   \delta_{[\ja\jb][\jc\jd]} \delta^{[\jc\jd][\je\jf]} &= \delta_{[\ja\jb]}^{[\je\jf]} \; ,
\end{align}
and the same is true for $\delta^{(\ja\jb)(\jc\jd)}$ and $\delta_{(\ja\jb)(\jc\jd)}$. Furthermore we have
\begin{align}
   \Omega^{\ja\jb} \delta_{[\ja\jc][\jb\jd]} &= \frac 5 4 \Omega_{\jc\jd} \; , &
   \delta^{[\jb\jc]}_{[\ja\jc]} &= \frac 5 4 \delta^\jb_\ja \; , &
   \delta^{[\ja\jb]}_{[\ja\jb]} &= 5 \; , \nonumber \\
   \Omega^{\ja\jb} \delta_{(\ja\jc)(\jb\jd)} &= \frac 5 2 \Omega_{\jc\jd} \; , &
   \delta^{(\jb\jc)}_{(\ja\jc)} &= \frac 5 2 \delta^\jb_\ja \; , &
   \delta^{(\ja\jb)}_{(\ja\jb)} &= 10 \; .
\end{align}   
We use the index pairs $(\ja\jb)$ and $[\ja\jb]$ as composite indices for the ${\bf 5}$ and ${\bf 10}$ representation;
they are raised and lowered using the above metrics and when having several of them we use the usual bracket notation 
for symmetrization and anti-symmetrization.

The following tensors represent some 
projections onto the irreducible components 
 of particular ${\rm USp}(4)$ representations:
\begin{align}
   \tau_{(\ja\jb)(\jc\jd)(\je\jf)} &= \Omega_{(\je(\ja} \Omega_{\jb)(\jc} \Omega_{\jd)\jf)} \; ,
         &&  \left[ ({\bf 10} \otimes {\bf 10})_{\text{asymm.}} \mapsto {\bf 10} \right] \;, \nonumber \\
   \tau_{(\ja\jb)[\jc\jd][\je\jf]} &= \Omega_{(\ja[\jc} \Omega_{\jd][\je} \Omega_{\jf]\jb)} \; , 
         &&   \left[ ({\bf 5} \otimes {\bf 5})_{\text{asymm.}} \mapsto {\bf 10} \right] \;,\nonumber \\
   \tau_{[\ja\jb](\jc\jd)(\je\jf)} &= \Omega_{[\ja(\jc} \Omega_{\jd)(\je} \Omega_{\jf)\jb]}  
                                      - \frac 1 4 \Omega_{\ja\jb} \delta_{(\jc\jd)(\je\jf)}  \; ,
         &&   \left[ ({\bf 10} \otimes {\bf 10})_{\text{symm.}} \mapsto {\bf 5} \right] \;,\nonumber \\
   \tau_{(\ja\jb)[\jc\jd][\je\jf][\jg\jh]} &= 
   \hat \tau_{(\ja\jb)\,\text{\bf [}[\jc\jd][\je\jf][\jg\jh]\text{\bf ]}}  \; ,
         &&   \left[ ({\bf 5} \otimes {\bf 5} \otimes {\bf 5})_{\text{asymm.}} \mapsto {\bf 10} \right]\;,
   \label{DefTaus}  
\end{align}
where
\begin{align}
   \hat \tau_{(\ja\jb)[\jc\jd][\je\jf][\jg\jh]} &= \Omega_{(\ja[\jc} \Omega_{\jd][\je} \Omega_{\jf][\jg} \Omega_{\jh]\jb)}
                                                   + \frac 1 4  \tau_{(\ja\jb)[\jc\jd][\je\jf]} \Omega_{\jg\jh}
                                                   \;.
\end{align}
The contractions of these $\tau$-tensors with $\Omega$ yield
\begin{align}
   \Omega^{\jd\jf} \tau_{(\ja\jb)(\jc\jd)(\je\jf)} &= - \frac 3 2 \delta_{(\ja\jb)(\jc\je)} \; , \nonumber  \\
   \Omega^{\jd\jf} \tau_{(\ja\jb)[\jc\jd][\je\jf]} &= \frac 1 2 \delta_{(\ja\jb)(\jc\je)} \; , &
   \Omega^{\jb\jd} \tau_{(\ja\jb)[\jc\jd][\je\jf]} &= - \delta_{[\ja\jc][\je\jf]} \; , \nonumber  \\
   \Omega^{\jd\jf} \tau_{[\ja\jb](\jc\jd)(\je\jf)} &= - \frac 3 2 \delta_{[\ja\jb][\jc\je]} \; , &
   \Omega^{\jb\jd} \tau_{[\ja\jb](\jc\jd)(\je\jf)} &= \frac 3 4 \delta_{(\ja\jc)(\je\jf)} \; , \nonumber  \\
   \Omega^{\jf\jh} \tau_{(\ja\jb)[\jc\jd][\je\jf][\jg\jh]} &= \frac 1 2 \tau_{[\jc\jd](\ja\jb)(\je\jf)} \; , &
   \Omega^{\jb\jd} \tau_{(\ja\jb)[\jc\jd][\je\jf][\jg\jh]} &= - \frac 3 4 \tau_{(\ja\jc)[\je\jf][\jg\jh]} \; , \nonumber  \\
   \Omega^{\jd\jg} \Omega^{\jf\jh} \tau_{(\ja\jb)[\jc\jd][\je\jf][\jg\jh]} &= \frac 3 8 \delta_{(\ja\jb)(\jc\je)} \; .
\end{align} 
Note that $\tau_{(\ja\jb)(\jc\jd)(\je\jf)}$ is totally antisymmetric in the three index pairs.
Since the ${\bf 10}$ is the adjoint representation, the structure constants of ${\rm USp}(4)$ are 
${\tau_{(\ja\jb)(\jc\jd)}}^{(\je\jf)}$. The ${\rm USp}(4)$ generators in the ${\bf 5}$ representation
are ${\tau_{(\ja\jb)[\jc\jd]}}^{[\je\jf]}$ and satisfy the algebra
\begin{align}
    {\tau_{(\ja\jb)[\je\jf]}}^{[\jg\jh]} {\tau_{(\jc\jd)[\jg\jh]}}^{[\ji\jj]}
  - {\tau_{(\jc\jd)[\je\jf]}}^{[\jg\jh]} {\tau_{(\ja\jb)[\jg\jh]}}^{[\ji\jj]}
              &= {\tau_{(\ja\jb)(\jc\jd)}}^{(\jg\jh)} {\tau_{(\jg\jh)[\je\jf]}}^{[\ji\jj]}  \; .
\end{align}
As defined above, $\tau_{(\ja\jb)[\jc\jd][\je\jf]}$ describes the mapping
$({\bf 5} \otimes {\bf 5})_{\text{asymm.}} \mapsto {\bf 10}$. However since
$({\bf 5} \otimes {\bf 5})_{\text{asymm.}} = {\bf 10}$ this must be a bijection. Indeed one finds
\begin{align} 
   x_{(\ja\jb)} &= \sqrt{2} \; {\tau_{(\ja\jb)}}^{[\jc\jd][\je\jf]} \, x_{[\jc\jd][\je\jf]} = \Omega^{\jc\jd} \, x_{[\ja\jc][\jb\jd]} 
   && \Leftrightarrow &
   x_{[\jc\jd][\je\jf]} &= \sqrt{2} \; {\tau^{(\ja\jb)}}_{[\jc\jd][\je\jf]} \, x_{(\ja\jb)} \; ,
   \label{SO5eqUSp4}
\end{align}
for tensors $x_{(\ja\jb)}$ and $x_{[\ja\jc][\jb\jd]}=-x_{[\jb\jd][\ja\jc]}$.
When regarding $(\bf{5} \otimes \bf{5})_{\text{asymm.}}$ as the adjoint representation of ${\rm SO}(5)$, 
formula \eqref{SO5eqUSp4} describes the isomorphism between the algebras of ${\rm USp}(4)$ and ${\rm SO}(5)$.
Some other useful relations in this context are
\begin{align}       
   \tau_{(\ja\jb)[\jc\jd][\je\jf]} \, {\tau^{(\ja\jb)}}_{[\jg\jh][\ji\jj]}
           &= \frac 1 2 \delta_{[\jc\jd]\,\text{\bf[}[\jg\jh]} \delta_{[\ji\jj]\text{\bf]}\,[\je\jf]} \; , &
   \Omega^{\jb\jd} \, \delta^{\phantom{[\jg}}_{[\je\jf]\,\text{\bf[}[\ja\jb]} \delta_{[\jc\jd]\text{\bf]}}^{[\jg\jh]}
                &= {\tau_{(\ja\jc)[\je\jf]}}^{[\jg\jh]} \; .
\end{align}
The last equation states that under the bijection~\eqref{SO5eqUSp4}
the generators of the ${\rm SO}(5)$ vector representation\footnote{
When denoting ${\rm SO}(5)$ vector indices by $\underline{\Ja}$, $\underline{\Jb}$, \ldots, \ 
the ${\rm SO}(5)$ generators 
in the vector representation 
are given by ${t_{\underline{\Ja\Jb},\underline{\Jc}}}^{\underline{\Jd}} = 
\delta^{\phantom{\Jd}}_{\underline{\Jc}[\underline{\Ja}} \delta_{\underline{\Jb}]}^{\underline{\Jd}}$.}
yield ${\tau_{(\ja\jb)[\jc\jd]}}^{[\je\jf]}$.

Also the five-dimensional $\epsilon$-tensor 
can be expressed in terms of $\Omega_{\ja\jb}$. A useful relation  is
\begin{align}
   \epsilon^{[\ja\jb][\jc\jd][\je\jf][\jg\jh][\ji\jj]} x_{[\jc\jd][\je\jf]} y_{[\jg\jh][\ji\jj]}
    &= 4 \tau^{[\ja\jb](\jc\jd)(\je\jf)} x_{(\jc\jd)} y_{(\je\jf)} \;,
\end{align}
where $x$ and $y$ in the $({\bf 5} \otimes {\bf 5})_{\text{asymm.}}={\bf 10}$
representation are related by \eqref{SO5eqUSp4}.

There is no singlet in the product of three ${\rm SO}(5)$ vectors and thus no invariant tensor of the
form $\tau_{[\ja\jb][\jc\jd][\je\jf]}$. This gives rise to the identity
\begin{align}        
   0 &= \delta^{[\ja}_{[\jc} \Omega_{\jd][\je} \delta_{\jf]}^{\jb]} - \text{traces} \nonumber \\
     &= \delta^{[\ja}_{[\jc} \Omega_{\jd][\je} \delta_{\jf]}^{\jb]}
     + \frac 1 4 \Omega^{\ja \jb} \eta_{[\jc \jd][\je \jf]} + \frac 1 4  \Omega_{\jc \jd} \delta^{[\ja \jb]}_{[\je \jf]}
     + \frac 1 4 \Omega_{\je \jf} \delta^{[\ja \jb]}_{[\jc \jd]} + \frac 1 {16} \Omega^{ij} \Omega_{kl} \Omega_{mn}  \; .
\end{align}
Using this equation one finds
\begin{align}
   \Omega^{\jc \jd} \lambda_{\jc [\ja} \mu_{\jb] \jd}  &= - \frac 1 4 \Omega_{\ja \jb} 
               \eta^{[\jc \jd][\je\jf]} \lambda_{[\jc\jd]}  \mu_{[\je\jf]}  \; , &
   \Omega^{\jc \jd} \lambda_{[\ja \jc]} \lambda_{[\jb \jd]} &= \frac 1 4 \Omega_{\ja \jb} 
               \eta^{[\jc \jd][\je\jf]} \lambda_{[\jc\jd]}  \lambda_{[\je\jf]}        
               \;,
\end{align}
for tensors $\lambda_{[\ja \jb]}$, $\mu_{[\ja \jb]}$ in the ${\bf 5}$ representation.

\section{$T$-tensor and quadratic constraints}
\setcounter{equation}{0}
\label{app:TQC}

In terms of the tensors~\eqref{DefTaus} defined in the previous section
the decomposition of the $T$-tensor into its ${\rm USp}(4)$ irreducible components
can be stated in the systematic form
\begin{align}
   {T_{(\ja\jb)[\jc\jd]}}^{[\je\jf]} &= \sqrt{2} \; \Omega^{\jg\jh} \, {X_{[\ja\jg][\jb\jh][\jc\jd]}}^{[\je\jf]} \nonumber \\
                                     &=
                                      - B {\tau_{(\ja\jb)[\jc\jd]}}^{[\je\jf]}
                                      - {B^{[\jg\jh]}}_{[\jc\jd]} {\tau_{(\ja\jb)[\jg\jh]}}^{[\je\jf]}
+ C^{[\jg\jh]} {\tau_{(\ja\jb)[\jg\jh][\jc\jd]}}^{[\je\jf]}
+ {C^{[\je\jf]}}_{(\jg\jh)} {\tau_{[\jc\jd](\ja\jb)}}^{(\jg\jh)} \;,
   \label{TTwithTaus}      
\end{align}
from which~\eqref{TABCD} is recovered with the explicit definitions of~\eqref{DefTaus}.
Similarly, the variation of the scalar potential under
$\delta_{\Sigma} {\cal V}_{M}{}^{ab}=\Sigma^{ab}{}_{cd}\,{\cal V}_{M}{}^{cd}$
takes the more concise form
\begin{align}
  \delta_{\Sigma} V &= 
          -\frac {g^{2}}{16} {B^{[\ja\jb]}}_{[\jc\jd]} 
          {B^{[\jc\jd]}}_{[\je\jf]} {\Sigma^{[\je\jf]}}_{[\ja\jb]}
+\frac {g^{2}}{32} B {B^{[\ja\jb]}}_{[\jc\jd]} {\Sigma^{[\jc\jd]}}_{[\ja\jb]}
          -\frac {g^{2}}{64} C^{[\ja\jb]} C_{[\jc\jd]} {\Sigma^{[\jc\jd]}}_{[\ja\jb]}
\nonumber \\[1ex]  &
{}
+\frac {g^{2}}{16} {C^{[\ja\jb]}}_{(\je\jf)} {C_{[\jc\jd]}}^{(\je\jf)} {\Sigma^{[\jc\jd]}}_{[\ja\jb]}
+\frac{g^{2}}8
{\tau^{(\ja\jb)}}_{[\je\jf][\ji\jj]} {\tau^{(\jc\jd)}}_{[\jg\jh][\jk\jl]} 
{C^{[\jg\jh]}}_{(\ja\jb)} C^{[\je\jf](\jc\jd)} 
\Sigma^{[\ji\jj][\jk\jl]} \;,
\end{align}
from which \eqref{VaryV} is deduced.

The quadratic constraint \eqref{quadratic} 
on the components $Y_{\Ja\Jb}$ and $Z^{\Ja\Jb,\Jc}$ of the embedding tensor $\Theta$
translates under the ${\rm USp}(4)$ split into quadratic constraints 
on the components
$B$, ${B^{\ja\jb}}_{\jc\jd}$, $C^{\ja\jb}$ and ${C^{\ja\jb}}_{\jc\jd}$ of the $T$-tensor. 
These constraints prove essential when checking the 
algebra of the supersymmetry transformation \eqref{SUSYrules}
and the invariance of the Lagrangian \eqref{L} under these transformations.
According to \eqref{sum} the quadratic constraint on $\Theta$ decomposes into 
a ${\bf \overline{5}}$, a ${\bf \overline{45}}$ and a ${\bf \overline{70}}$ under ${\rm SL}(5)$
which under ${\rm USp}(4)$ branch as
\begin{align}
   {\bf\overline 5} ~\rightarrow~ & {\bf 5} \, , &
   {\bf\overline{45}} ~\rightarrow~& {\bf10} \oplus {\bf 35} \, , &
   {\bf\overline{70}} ~\rightarrow~& {\bf 5} \oplus {\bf30} \oplus {\bf35} \, ,
\end{align}
In closed form, these constraints have been given in~\eqref{QZB2}. 
The check of supersymmetry of the Lagrangian however needs the explicit 
expansion of these equations in terms of $B$  and $C$.
The two ${\bf 5}$ parts and the ${\bf 10}$ part read
\begin{align}
     4 B C^{[\ja\jb]} 
       - {B^{[\ja\jb]}}_{[\jc\jd]}  C^{[\jc\jd]} 
       - 4 {B^{[\ji\jj]}}_{[\jc\jd]} {C^{[\jc\jd]}}_{(\jg\jh)} {\tau^{(\jg\jh)[\ja\jb]}}_{[\ji\jj]}  &= 0 \; , 
       \nonumber \\
     B B^{[\ja\jb]} 
       + {B^{[\ja\jb]}}_{[\jc\jd]}  C^{[\jc\jd]} 
        + \tau^{[\ja\jb](\jc\jd)(\je\jf)} {C^{[\jg\jh]}}_{(\jc\jd)} C_{[\jg\jh](\je\jf)} &= 0 \; , 
\nonumber \\
     {\tau_{[\jc\jd]}}^{(\ja\jb)(\jg\jh)} {B^{[\jc\jd]}}_{[\je\jf]} {C^{[\je\jf]}}_{(\jg\jh)} &= 0 \;,
\label{Con5}
\end{align}      
respectively.
In particular, a proper linear combination of the first two equations yields the quadratic
relation~\eqref{A1A1A2A2} cited in the main text.
The two ${\bf 35}$ parts of the quadratic constraint are
\begin{align}
     {\tau_{(\jc\jd)}}^{[\ja\jb][\je\jf]} B C_{[\je\jf]} + B {C^{[\ja\jb]}}_{(\jc\jd)}
     + {\tau_{(\jc\jd)}}^{[\je\jf][\jg\jh]} {B^{[\ja\jb]}}_{[\je\jf]} C_{[\jg\jh]}
     + {B^{[\ja\jb]}}_{[\je\jf]} {C^{[\je\jf]}}_{(\jg\jh)}  &= 0  \; , 
   \nonumber \\[1.5ex]
   \mathbb{P}_{{\bf 35}} \Big( 
     B {C^{[\ja\jb]}}_{(\jc\jd)} 
       - 4 \tau^{(\je\jf)[\jg\jh][\ja\jb]} \tau_{(\jc\jd)[\ji\jj][\jk\jl]} {C^{[\ji\jj]}}_{(\je\jf)} {B^{[\jk\jl]}}_{[\jg\jh]} 
       \qquad \qquad \qquad \qquad \phantom{a} \nonumber & \\
       - 3 {\tau_{[\je\jf](\jc\jd)}}^{(\jg\jh)} C^{[\je\jf]} {C^{[\ja\jb]}}_{(\jg\jh)}
    + 4  {\tau^{(\je\jf)[\ja\jb]}}_{[\jg\jh]} {\tau_{[\ji\jj](\jc\jd)}}^{(\jk\jl)} {C^{[\ji\jj]}}_{(\je\jf)} {C^{[\jg\jh]}}_{(\jk\jl)} 
     \Big)
       &= 0 \; ,
   \label{Con35}    
\end{align}
where the projector $\mathbb{P}_{\bf 35}$ is defined by
\begin{align}
   \mathbb{P}_{{\bf 35}}\left( {X^{[\ja\jb]}}_{(\jc\jd)} \right) &=
    \left( \delta^{[\ja\jb]}_{[\je\jf]} \delta_{(\jc\jd)}^{(\jg\jh)} 
      - {\tau_{(\jc\jd)}}^{[\ja\jb][\ji\jj]} {\tau^{(\jg\jh)}}_{[\je\jf][\ji\jj]}
      - \frac 4 3 {\tau^{[\ja\jb]}}_{(\jc\jd)(\ji\jj)} {\tau_{[\je\jf]}}^{(\jg\jh)(\ji\jj)} \right) {X^{[\je\jf]}}_{(\jg\jh)} \; .
\end{align}
Note that also the first equation of \eqref{Con35} has to be projected with $\mathbb{P}_{\bf 35}$
in order to reduce it to a single irreducible part. However this equation is satisfied 
also without the projection,
since it contains the above ${\bf 10}$ and one of the ${\bf 5}$ constraints as well.\footnote{
Indeed this first equation of \eqref{Con35} is equivalent to~\eqref{QZB1}.}
Finally the ${\bf 30}$ component of the quadratic constraint is 
obtained by completely symmetrizing \eqref{QZB2}
in the three free index pairs, i.e.
\begin{align}
   Z^{(\jg\jh)\text{\bf(}[\ja\jb]} {T_{(\jg\jh)}}^{[\jc\jd][\je\jf]\text{\bf)}} &= 0 \;.
\end{align}
\end{appendix}


\providecommand{\href}[2]{#2}\begingroup\raggedright\endgroup

\end{document}